\documentclass[twocolumn]{aastex631} % see doc for all arg options

\usepackage{graphicx}
\usepackage{amsmath} 
\usepackage{xspace}
\usepackage{mathtools}
\newcommand{\Tierras}{\textit{Tierras}\xspace}

\newcommand{\PAIRITEL}{\textit{PAIRITEL}\xspace}

\newcommand{\TWOMASS}{\textit{2MASS}\xspace}

\definecolor{bmaroonclr}{RGB}{140,0,26}

\newcommand\bedit[1]{\textcolor{black}{#1}}

\defcitealias{GaMe20}{GM20}
\defcitealias{GaMe23}{GM23}

\shorttitle{The \Tierras Observatory}
\shortauthors{Garc\'ia-Mej\'ia et al.}
\graphicspath{{./}{figures/}}

\begin{document}

\title{The \Tierras Observatory: \\ Ultra-Precise Photometry for Characterizing Exoplanets and Their Host Stars}

\author[0000-0003-1361-985X]{Juliana Garc\'ia-Mej\'ia}
\altaffiliation{51 Pegasi B Fellow, MIT Pappalardo Physics Fellow}
\affiliation{Kavli Institute for Astrophysics and Space Research, Massachusetts Institute of Technology, Cambridge, MA 02139, USA}
\affiliation{Center for Astrophysics \textbar\ Harvard \& Smithsonian, 60 Garden Street, Cambridge, MA 02138, USA}

\author[0000-0003-2171-5083]{Patrick Tamburo}
\affiliation{Center for Astrophysics \textbar\ Harvard \& Smithsonian, 60 Garden Street, Cambridge, MA 02138, USA}

\author[0000-0002-9003-484X]{David Charbonneau}
\affiliation{Center for Astrophysics \textbar\ Harvard \& Smithsonian, 60 Garden Street, Cambridge, MA 02138, USA}

\author[0000-0002-1311-4942]{Daniel Fabricant}
\affiliation{Center for Astrophysics \textbar\ Harvard \& Smithsonian, 60 Garden Street, Cambridge, MA 02138, USA}

\author{Jonathan M. Irwin}
\affiliation{Institute of Astronomy, University of Cambridge, Madingley Road, Cambridge CB3 0HA, UK}

\author{Robert Fata}
\affiliation{Center for Astrophysics \textbar\ Harvard \& Smithsonian, 60 Garden Street, Cambridge, MA 02138, USA}

\author{Joseph M. Zajac}
\affiliation{Center for Astrophysics \textbar\ Harvard \& Smithsonian, 60 Garden Street, Cambridge, MA 02138, USA}

\author{Peter E. Doherty}
\affiliation{Center for Astrophysics \textbar\ Harvard \& Smithsonian, 60 Garden Street, Cambridge, MA 02138, USA}

\author[0000-0002-2924-2893]{Brian McLeod}
\affiliation{Center for Astrophysics \textbar\ Harvard \& Smithsonian, 60 Garden Street, Cambridge, MA 02138, USA}

\author[0000-0001-8963-4437]{Pascal Fortin}
\affiliation{Fred Lawrence Whipple Observatory, Smithsonian Astrophysical Observatory, 670 Mount Hopkins Road, Amado, AZ 85645, USA}

\author[0000-0002-2830-5661]{Michael L. Calkins}
\affiliation{Fred Lawrence Whipple Observatory, Smithsonian Astrophysical Observatory, 670 Mount Hopkins Road, Amado, AZ 85645, USA}

\author[0000-0002-9789-5474]{Gilbert A. Esquerdo}
\affiliation{Center for Astrophysics \textbar\ Harvard \& Smithsonian, 60 Garden Street, Cambridge, MA 02138, USA}
\affiliation{Planetary Science Institute, 1700 East Fort Lowell Road, Suite 106, Tucson, AZ 85719, USA}

%%%%%%%%%%%%%%%%%%%%%%%%%%%%%%%%%%%%%%%%%%%%%%%%%%%%%%%%%%%%%%%%%%%%%%%%%%%%
\begin{abstract}

The \Tierras Observatory is a fully robotic 1.3-m telescope at the Fred Lawrence Whipple Observatory in southern Arizona. \textit{Tierras} achieves sub-part-per-thousand (ppt) photometric precision on two timescales: intra-night, to detect and  characterize transits of planets around nearby mid-to-late M dwarfs, and night-to-night across an observing season, to characterize the long-term variability of their host stars.  We describe the facility and instrument design, our photometric pipeline, and our on-sky performance. We present the first ground-based light curves with sub-ppt precision over timescales of months. This precision enables a range of exoplanet-related science, including the measurement of host star rotation periods, the detection of long-period transits, and the discovery of additional terrestrial planets in known M dwarf systems.

\end{abstract}

\section{Introduction}
\label{sec:intro}

Although the study of exoplanets has seen dramatic advances in the past decades, analogs of many Solar System objects remain beyond our reach. Most known terrestrial exoplanets are significantly larger than Earth, no exomoon or exoring systems have been discovered, and we have only begun to characterize the long-term variability of low-mass stars and its impact on attendant worlds. Time-series photometry can address these questions, but two obstacles limit the achievable precision of ground-based observatories. 1) systematic errors from time-variable precipitable water vapor (PWV) that induce photometric variability exceeding 1\% on transit-like timescales in broadband photometry \citep{St07, Bl08, Bl11, Murray2020, Tamburo2022a}. 2) a long-standing night-to-night precision ceiling of $\sim$1~part-per-thousand (ppt; 0.1\%)\footnote{This limit is often defined as 0.001~mag or 1~mmag, which is $\approx$1~ppt in normalized flux.} that has persisted for decades \citep{Lockwood2007}, with ground-based efforts typically achieving precisions closer to 3--5~ppt \citep{Radick1983, Young1991}.

In 2018, we began work on the \Tierras Observatory, a fully robotic 1.3-m facility at the Fred Lawrence Whipple Observatory in southern Arizona, designed to overcome both of these limitations. \Tierras' ultra-precise photometer, described in \citet{GaMe20} and \citet{GaMe23} (hereafter \citetalias{GaMe20} and \citetalias{GaMe23}), is based on a custom narrow-band near-infrared (NIR) filter that selects a spectral window largely free of telluric water features. Here, we describe the facility, the photometric pipeline, the on-sky performance, and our current science program.

\subsection{Science Goals}\label{sec:science}
\Tierras science requires two complementary types of photometric precision.  1) \textit{Intra-night} precision on timescales of tens of minutes to several hours, relevant to transits and eclipses.  2) \textit{Night-to-night} precision, the stability of nightly mean flux measurements across weeks to months, relevant to stellar variability and long-baseline monitoring.

\subsubsection{Characterizing known exoplanet host stars.} 

Transiting terrestrial planets around mid-to-late M dwarfs are currently the only Earth-like planets whose atmospheres can potentially be characterized with JWST \citep{Morley2017, Pa25}. More than a dozen such systems have been found within 15~pc,\footnote{Query of the NASA Exoplanet Archive (\url{https://exoplanetarchive.ipac.caltech.edu}) on 2026 January 26 with criteria: $R_p < 1.5~R_\oplus$, $d < 15$~pc, $T_{\rm eff} < 4000$~K, $R_\star < 0.45~R_\odot$. TRAPPIST-1 added manually as its distance is not listed in the default parameter set.} including TRAPPIST-1 \citep{Gi17}, LHS~1140 \citep{Di17}, and L~98-59 \citep{Ko19}, but no confident detection of a temperate terrestrial exoplanet atmosphere exists \citep[e.g.,][]{Greene2023, Zi23, Cadieux2024, Bello-Arufe2025, Rochon2026}. Interpreting JWST transmission spectra of these planets requires understanding the host star's magnetic activity, as unocculted spots and faculae may contaminate the observations at the level of hundreds of ppm \citep{Lim23, Zi23}. Measuring the stellar rotation period constrains this contamination and, via gyrochronology \citep{Barnes2003, Barnes2007}, the system age. Rotation periods also deliver 3D obliquities when combined with Rossiter-McLaughlin measurements \citep{Masuda2020, Albrecht2022}, constraining planetary formation histories. We are pursuing rotation periods for a sample of exoplanet host stars with \Tierras, with results published for TOI-2364~b \citep{Tamburo2025a}, TOI-2374~b \citep{Yee2025}, and TOI-3884~b \citep{Tamburo2025b}.

\subsubsection{Rotation census of nearby low-mass stars.} 

Ground-based monitoring from MEarth \citep{Ne16, Ne18} combined with TESS photometry \citep{Me20, Me22} has yielded rotation periods for roughly half the mid-to-late M dwarfs within the volume-complete 15~pc sample \citep{Wi21, Pass2023, Winters2026}. The other half, likely slow rotators with $P_\text{rot}$ in the range of 50--150~days, lack rotation periods due to TESS's 27-day baseline and MEarth's atmospheric systematics that prevent detection of low-amplitude variability. Completing this census is critical: fully convective M dwarfs have rotation and activity histories that produce cumulative XUV fluences 2--3 times larger than canonical scaling relations \citep{Pa25}, worsening prospects that their rocky planets retain an atmosphere. We are using \Tierras to measure the missing rotation periods and to extend long-term monitoring into the substellar regime for the first time, targeting nearby L dwarfs (Tamburo et al.\ in prep.).

\subsubsection{Long-period and missed planets, moons and rings.} 

Kepler and TESS have discovered hundreds of transiting exoplanets with $P > 100$~days \citep{Christiansen2025}, but characterizing their orbits requires precise ground-based follow-up to refine ephemerides and constrain transit timing variations. With its night-to-night stability, \Tierras can detect transits of long-period planets without observing ingress or egress, as demonstrated for HIP~41378~f \citep{GarciaMejia2026}. We are observing M~dwarfs with known transiting planets to find additional planets undetectable by TESS \citep[e.g.][]{Ba19,De22} and targeting known transiting systems to conduct systematic ground-based searches for exomoons and circumplanetary rings.

All of this science benefits from high photometric precision, not just across single nights but across observing seasons. The \Tierras design was driven by the goal of achieving 0.25~ppt stability on intra-night and night-to night timescales, allowing detection of an Earth sized planet transiting a mid-M dwarf at $3\sigma$ significance and measurement of the low-amplitude rotational variability of slowly rotating, low-mass stars.

\subsection{The Precipitable Water Vapor Problem}\label{sec:pwv}
For ground-based differential photometry, achieving 0.25~ppt precision is challenging. Second-order differential extinction arises from the color-dependence of molecular absorption between red target stars and bluer comparison stars with variable atmospheric water vapor. Because the target and comparison stars have different spectral energy distributions, water vapor column changes affect them unequally, inducing spurious photometric variations that can exceed 1\% \citep{St07, Bl08, Bl11, Murray2020, Tamburo2022a}. This effect is most severe for long-pass filters at NIR wavelengths, where strong telluric water absorption bands are prevalent.

Observing a mid-M dwarf with a 4600~K reference star through the 750~nm long-pass bandpass of MEarth \citep{Ir15}, \citet{Ba17} estimate a differential photometric bias of $\sim$2~ppt per mm of PWV in dry conditions ($<5$~mm), and note that the effect grows for hotter reference stars. In practice, MEarth light curves show night-to-night fluctuations as large as 1\% correlated with ground-level humidity \citep{Be12}. TRAPPIST \citep{Gi11} and SPECULOOS \citep{De18} use a 720~nm $I+z'$ filter and face similar systematics; \citet{Pe23} showed this bandpass is highly sensitive to PWV at Paranal, with changes $>$1~ppt observed 37\% of the time on transit-like timescales. All three facilities achieve typical intra-night precision of $\sim$1~ppt and night-to-night precision exceeding 3~ppt. Space-based observatories have different constraints: TESS achieves $\sim$0.06~ppt on phase-folded data of bright targets, but $\sim$1~ppt for single transits of faint M dwarfs due to its small aperture, while CHEOPS \citep{Be21} offers excellent precision but limited availability and poorer precision for late-type targets. We describe how \Tierras performance compares below.

This paper is organized as follows. In Section~\ref{sec:facility}, we describe the \Tierras telescope, instrument components, and robotic control system. Section~\ref{sec:commissioning} summarizes the commissioning process and subsequent upgrades. Section~\ref{sec:pipeline} details the photometric pipeline, with a particular focus on our reference star weighting scheme. Section~\ref{sec:performance} presents the on-sky performance on single nights and across many nights, along with a summary of observing statistics since first science light in 2022. We conclude in Section~\ref{sec:summary} with a summary and outlook for \Tierras.

\section{Facility Description} \label{sec:facility}

\subsection{Site and Telescope} \label{sec:site}

\begin{figure*}[ht]
\begin{center}
\includegraphics[width=\linewidth]{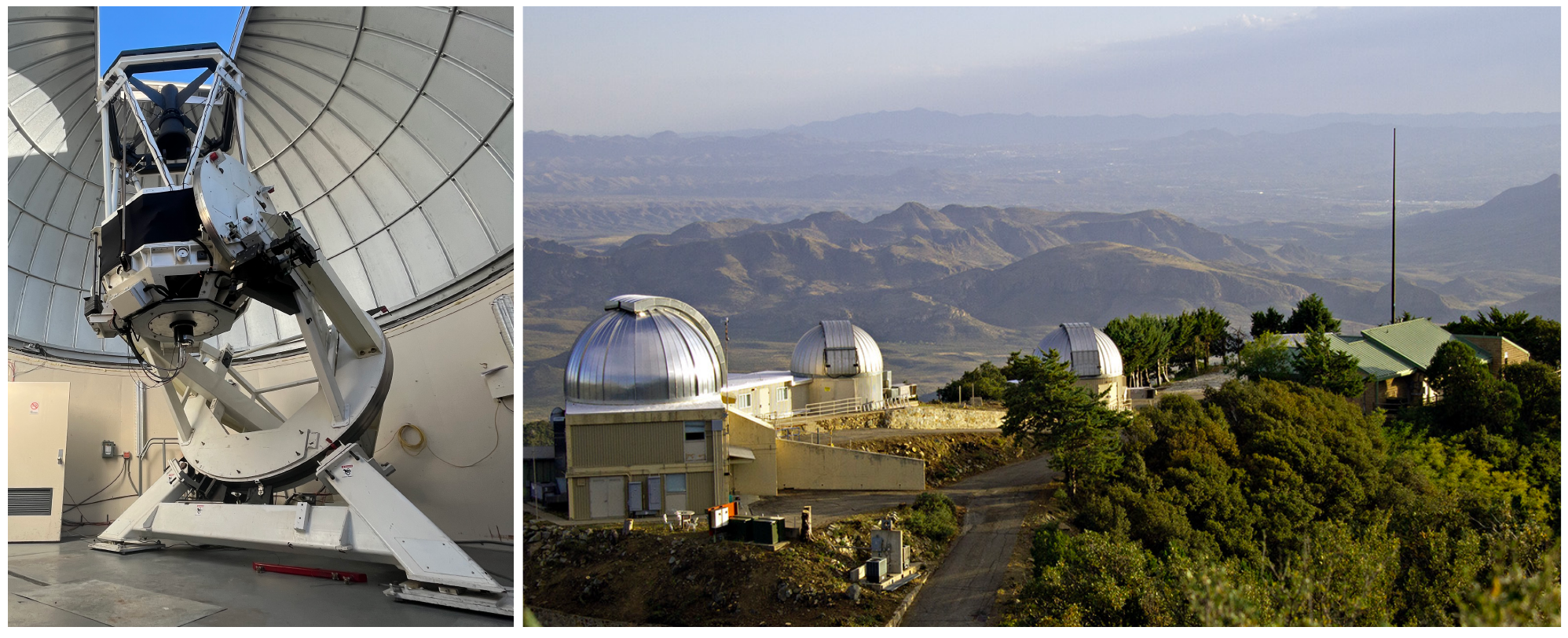}
\end{center}
\caption{\label{fig:1.3m} \textit{Left:} The 1.3-m telescope pictured inside its dome with the fully installed \Tierras instrument, ca.\ 2024; compare to \citetalias{GaMe20} where the telescope is shown prior to instrument installation. \textit{Right:} The FLWO ridge. \textit{From left to right}: the Tillinghast 1.5-m telescope (hosting the TRES and FAST spectrographs, \citealt{Sz07,Fa98b}), the 1.2-m telescope (hosting KeplerCam, \citealt{Sz05}), and the 1.3-m \Tierras Observatory. MINERVA \citep{Wil19b}, HAT-5 \citep{Ba18}, and MEarth \citep{Ir15} are also located on the ridge but are not shown in the frame.}
\end{figure*}

\Tierras uses a refurbished 1.3-m Cassegrain equatorial telescope at the Fred Lawrence Whipple Observatory (FLWO), located at an elevation of 2306~m on a ridge below the summit of Mt.\ Hopkins, Arizona (N $31\arcdeg 40\arcmin 59.61\arcsec$, W $110\arcdeg 52\arcmin 31.14\arcsec$; Figure~\ref{fig:1.3m}). Both mirrors are Corning ultra-low expansion glass, providing a native focal ratio of $f/13.5$ and plate scale of $11.94\arcsec$~mm$^{-1}$; full optical specifications are listed in Table~\ref{tab:Telpresc}.

M3 Engineering and Technology Corp built the telescope between 1995 and 1996, with mirrors polished by Rayleigh Optical Corporation. The telescope was commissioned in 1997 and used by the Two Micron All Sky Survey \citep[\TWOMASS;][]{Sk06} to scan 92\% of the northern hemisphere between 1997 June and 2001 February.  The telescope was recommissioned as the Peters Automated Infrared Imaging Telescope \citep[\PAIRITEL;][]{Bl06}, a robotic gamma-ray burst follow-up facility that operated from 2003 to 2012. The facility was dormant until 2018, when a feasibility study confirmed that the telescope mount, dome, and mirrors had been sufficiently preserved to serve as the platform for \Tierras. Refurbishing the existing facility, owned by the Center for Astrophysics (CfA), was at least $4\times$ cheaper than constructing a new telescope.

We collaborated with DFM Engineering\footnote{\url{https://www.dfmengineering.com/}} to carry out the refurbishment on-site in 2020 January. We replaced the original telescope control system with DFM's TCSGalil, upgraded the incremental position encoders to Renishaw 26-bit absolute encoders with new DC servomotors on both axes, and refurbished the secondary mirror focus stage. New azimuth motors and a position encoder were installed on the dome, along with a radio-controlled shutter, enabling fully robotic observing. A full account of the refurbishment is provided in \citetalias{GaMe20} and \citetalias{GaMe23}.

% Figure: The 1.3m Telescope

% Table: System Parameters
\begin{table}[ht!]
\centering
\caption{System parameters for the 1.3-m telescope. Columns list parameter symbol (where applicable), description, and value. Detailed specifications for each \Tierras instrument component are provided in Table~\ref{tab:tierras_instrument}.}
\label{tab:Telpresc}
{\scriptsize
\begin{tabular}{l@{\hspace{4pt}}l@{\hspace{4pt}}l}
\hline
\hline
\multicolumn{3}{c}{\textbf{Site}} \\
\hline
Coordinates & N $31\arcdeg 40\arcmin 59.61\arcsec$, & \\
 & W $110\arcdeg 52 \arcmin 31.14\arcsec$ & \\
Altitude & \multicolumn{2}{l}{2306 m} \\
\hline
\multicolumn{3}{c}{\textbf{Primary Mirror (M1)}} \\
\hline
$D_{\rm M1}$ & Mechanical diameter & 1300 mm \\
$D_{\rm M1, clear}$ & Clear diameter & 1280 mm \\
$D_{\rm M1, hole}$ & Hole diameter & 200 mm \\
$K_{\rm M1}$ & Conic constant & $-1.000$ \\
& & (parabolic) \\
$R_{\rm M1}$ & Radius of curvature & 5200 mm \\
Material & Corning ULE glass & \\
\hline
\multicolumn{3}{c}{\textbf{Secondary Mirror (M2)}} \\
\hline
$D_{\rm M2}$ & Mechanical diameter & 232 mm \\
$D_{\rm M2, clear}$ & Clear diameter & 228 mm \\
$K_{\rm M2}$ & Conic constant & $-1.847$ \\
& & (hyperbolic) \\
$R_{\rm M2}$ & Radius of curvature & 965.7 mm \\
Material & Corning ULE glass &  \\
\hline
\multicolumn{3}{c}{\textbf{Native Cassegrain Focus}} \\
\hline
$f/\#$ & Focal ratio & 13.5 \\
$S$ & Plate scale & $11.94\arcsec$ mm$^{-1}$ \\
$z_{\rm img}$ & M1 vertex to focal plane & 527 mm \\
\hline
\multicolumn{3}{c}{\textbf{\textit{2MASS N/SWIR} Camera (1997--2012)}} \\
\hline
$f/\#$ & Focal ratio (SWIR) & 3.53 \\
$S$ & Plate scale & $1.99\arcsec$ pix$^{-1}$ \\
FOV & Field of view & $0.14^{\circ}$ sq. \\
Filters & $J$, $H$, $K_s$ & \\
Detectors & $256 \times 256$ NICMOS3 (HgCdTe), &\\
 & one per channel (three total) & \\
& Pixel size & 40 $\mu$m \\
Obs.\ mode & Dec.\ scanning at $57\arcsec$ s$^{-1}$; &  \\
 & eff.\ FOV $8.5\arcmin \times 6\arcdeg$ per 7.8 s & \\
\hline
\multicolumn{3}{c}{\textbf{\Tierras Instrument (2021--)}} \\
\hline
$f/\#$ & Focal ratio & 5.6 \\
$S$ & Plate scale & $0.43\arcsec$ pix$^{-1}$ \\
FOV & Field of view & $0.48^{\circ} \times 0.24^{\circ}$ \\
$z_{\rm img}$ & M1 vertex to focal plane & 288.98 mm \\
Obs.\ mode & Staring at one field; & \\
 & usually 1--60~s exposures & \\
\hline
\end{tabular}}
\end{table}

\subsection{Instrument Components} \label{sec:instrument}

The \Tierras instrument components target each major noise source in the photometric error budget. At the design phase, the dominant contributions to the root-mean-square error for a ground-based differential measurement are estimated as
\begin{equation}
\sigma_{\rm tot} = \sqrt{\sigma_{\rm phot}^2 + \sigma_{\rm scint}^2 + \sigma_{\rm PWV}^2 + \sigma_{\rm sky}^2}
\label{eq:error_budget}
\end{equation}
where $\sigma_{\rm phot}$ is the photon noise from the target and reference stars, $\sigma_{\rm scint}$ is the scintillation noise for an ensemble of uncorrelated comparison stars, $\sigma_{\rm PWV}$ accounts for the second-order differential extinction discussed in Section~\ref{sec:pwv}, and $\sigma_{\rm sky}$ is the contribution from sky background. \Tierras was designed to limit $\sigma_{\rm PWV}$ to $\lesssim$0.25~ppt and stochastic errors ($\sigma_{\rm phot}$, $\sigma_{\rm scint}$) to below the transit depth of an Earth-sized planet around a $0.3~R_\odot$ star within 15~pc. 

Each term in Equation~\ref{eq:error_budget} motivated a specific instrument design choice, detailed in the following sections: a custom narrow-band filter ($\sigma_{\rm PWV}$), a four-lens focal reducer and field flattener ($\sigma_{\rm scint}$), a deep-depletion CCD operated in frame transfer mode ($\sigma_{\rm phot}$), and custom baffles ($\sigma_{\rm sky}$). Table~\ref{tab:tierras_instrument} summarizes their key specifications; design details are provided in \citetalias{GaMe20} and \citetalias{GaMe23}. 

\subsubsection{Custom Bandpass Filter}\label{subsec:filter}

Our filter design selects a spectral region free of strong telluric water features, avoiding the need for \textit{post hoc} correction. To determine the optimal filter parameters, we simulated the flux of target and comparison stars through the complete optical system. The stellar flux (in ergs~s$^{-1}$~cm$^{-2}$) through the $n$th filter candidate is:\footnote{We retain the prime notation from \citetalias{GaMe23} for consistency.}
\begin{equation}
F'_{X,n} = \int_0^\infty f'_X(\lambda) \, T'_n(\lambda, \mathrm{PWV}) \, d\lambda
\label{eq:flux}
\end{equation}
where $f'_X(\lambda)$ is a calibrated stellar spectrum (in ergs~s$^{-1}$~cm$^{-2}$~\AA$^{-1}$) from the HST CALSPEC catalog \citep{Bo14b}. We used Gliese~555 (M3.5V) and vB8 (M7V) as representative targets spanning the mid-to-late M dwarf range, with 18~Sco (G2V) as the comparison star. The total system transmission $T'_n(\lambda, \mathrm{PWV})$ is the product of four wavelength-dependent terms:
\begin{eqnarray}
T'_n(\lambda, \mathrm{PWV}) & = & T_{\mathrm{CCD}}(\lambda) \times T'_{\mathrm{Tel}}(\lambda) \times T'_{\mathrm{Filt},n}(\lambda) \nonumber \\
& & \times \; W(\lambda, \mathrm{PWV})
\label{eq:transmission}
\end{eqnarray}
where $T_{\mathrm{CCD}}(\lambda)$ is the CCD QE, $T'_{\mathrm{Tel}}(\lambda)$ is the optical transmission fraction due to the telescope mirror reflectivities and auxiliary optics transmissivities, $T'_{\mathrm{Filt},n}(\lambda)$ is the filter transmission function for the $n$th candidate, and $W(\lambda, \mathrm{PWV})$ is the telluric water vapor absorption spectrum, scaled from a reference TAPAS spectrum \citep{Be14}. Following the convention in \citet{Ba17}:
\begin{equation}
W(\lambda, \mathrm{PWV}) = W(\lambda, \mathrm{PWV}_0)^{(\mathrm{PWV}/\mathrm{PWV}_0)}
\label{eq:telluric}
\end{equation}
where $\mathrm{PWV}_0 = 9.592$~mm. We quantified PWV error through the $n$th filter as the fractional change in the target-to-comparison flux ratio between dry (0~mm) and wet (12~mm) atmospheric conditions:
\begin{equation}
\sigma_{\mathrm{PWV},n} = \left| 1 - \frac{(F'_{1,n}/F'_{2,n})_{\mathrm{PWV}=0\,\mathrm{mm}}}{(F'_{1,n}/F'_{2,n})_{\mathrm{PWV}=12\,\mathrm{mm}}} \right|
\label{eq:pwv}
\end{equation}
where $F'_{1,n}$ and $F'_{2,n}$ are the fluxes of the target and comparison stars, respectively. An exhaustive description of the parameter space search is provided in \citetalias{GaMe23}.

\begin{figure*}[ht]
    \centering
    \includegraphics[width=\textwidth]{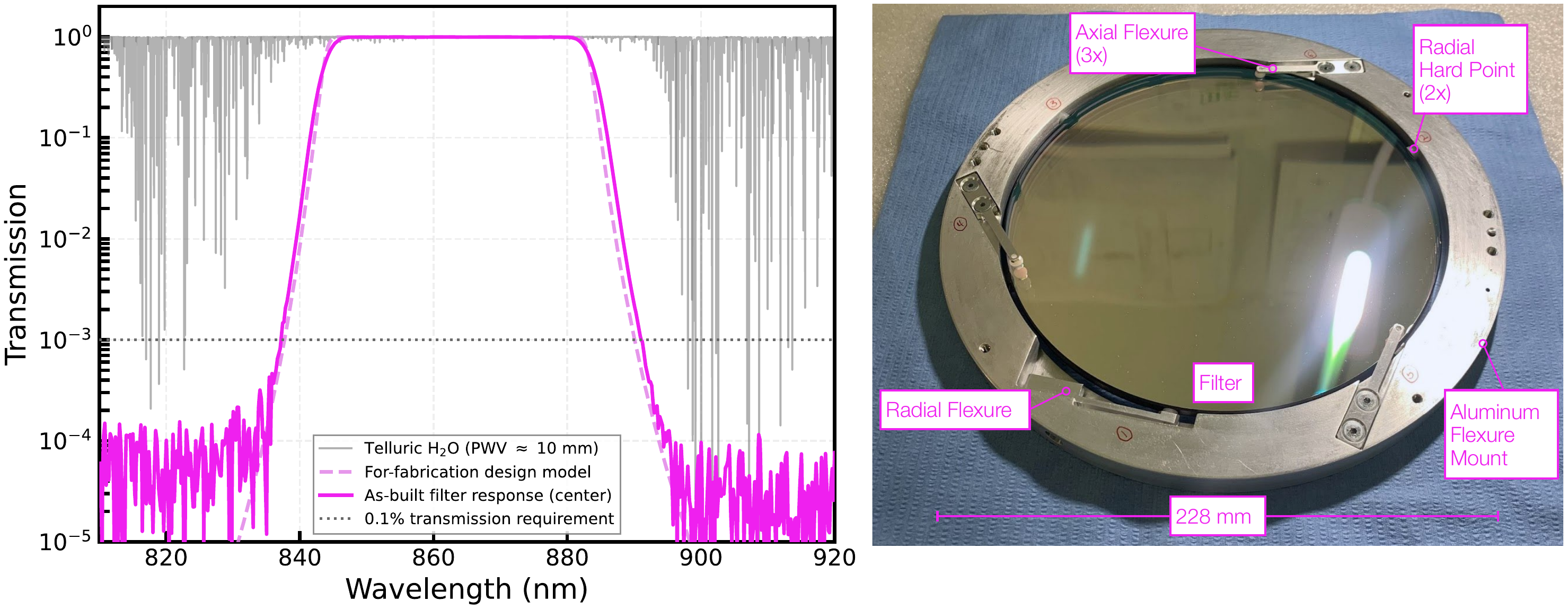}
    \caption{\textit{Left:} Transmission of the \Tierras\ custom bandpass filter as a function of wavelength, plotted on a logarithmic scale. The as-built filter response (solid pink), measured by Asahi Spectra Co. at the center of the filter substrate, is shown alongside the for-fabrication design model (dashed pink). The filter is centered at $\lambda_C \approx 863.5$~nm with a FWHM of $40.2$~nm, and achieves out-of-band attenuation of $T \leq 0.1\%$ for $300$--$834$~nm and $893$--$1100$~nm (dotted line). The telluric H$_2$O spectrum (gray) is shown for reference, scaled to a precipitable water vapor level of $9.592$~mm. \textit{Right:} The \Tierras\ bandpass filter ($180.3$~mm, fused silica) installed in its aluminum flexure mount ($228$~mm outer diameter). The three axial flexures and one radial flexure that kinematically support the filter substrate are labeled, along with the two radial hard points that constrain lateral motion. The scale bar indicates the outer diameter of the bezel.}
    \label{fig:filter_mosaic}
\end{figure*}

The search revealed that PWV performance is insensitive to filter slope shape but critically dependent on the locations of $\lambda_B$ and $\lambda_R$: the blue-ward and red-ward wavelengths beyond which strong water absorption features begin. Achieving attenuation $T \leq 0.1\%$ for $\lambda \leq \lambda_B = 834$~nm and $\lambda \geq \lambda_R = 893$~nm is the single most important specification, as these wavelengths bracket strong telluric water features while enclosing a relatively clean atmospheric window where M dwarfs emit copiously (Figure~\ref{fig:filter_mosaic}). We collaborated closely with the manufacturer, Asahi Spectra Co., Ltd.\footnote{\url{https://www.asahi-spectra.com/}}, to determine which combinations of center wavelength, bandwidth, and slope sharpness were achievable while guaranteeing our attenuation requirements across the full 174.41~mm clear aperture for all incidence angles ($0$--$4^\circ$) in our $f/13.5$ beam. The resulting filter is an eight-Fabry-Perot-cavity bandpass design deposited via ion beam sputtering onto a fused-silica substrate, enabling exceptionally steep slopes: the as-built filter achieves $\Delta\lambda_{90} \geq 0.85 \times \mathrm{FWHM}$ and $\Delta\lambda_{10} \leq 1.15 \times \mathrm{FWHM}$, where $\Delta\lambda_{90}$ and $\Delta\lambda_{10}$ are the bandwidths measured at 90\% and 10\% of peak transmission, respectively.

The as-built filter is centered at $\lambda_C = 863.5$~nm with a FWHM of 40.2~nm, achieves peak transmission exceeding 99\%, and meets the out-of-band attenuation specification at all five test locations across its surface. The filter limits PWV-induced photometric errors to $\lesssim 0.25$~ppt even for an extreme PWV change of 12~mm, an improvement of more than an order of magnitude over the long-pass filters used by MEarth \citep{Ir15} and SPECULOOS \citep{Pe23}; Table~\ref{tab:pwv} summarizes the predicted PWV error ceiling and representative photon noise for M3.5V and M7V targets. The filter is housed in a three-point kinematically-balanced aluminum flexure mount, chosen to facilitate maintenance and avoid potential long-term chemical interactions between RTV silicone rubber and the ion-sputtered coatings (Figure~\ref{fig:filter_mosaic}).

\begin{table}
\centering
\caption{Predicted PWV error ceiling and representative photon noise through the \Tierras filter for M dwarf targets at 0$^\circ$ AOI. The PWV error is calculated for an extreme 0--12~mm PWV swing; photon noise assumes a G2V comparison star, targets at 15~pc, and a 60-second exposure.}
\label{tab:pwv}
\begin{tabular}{lcc}
\hline
\hline
Target Star & $\sigma_{\mathrm{PWV}}$ (ppm) & $\sigma_{\mathrm{phot}}$ (ppm) \\
\hline
Gliese 555 (M3.5V) & 138 & 392 \\
vB8 (M7V) & 230 & 1999 \\
\hline
\end{tabular}
\end{table}

\subsubsection{Focal Reducer and Field Flattener}

Precise differential photometry requires monitoring comparison stars to correct for atmospheric transparency variations. \citet{Nu08} showed that a field of $\sim$0.5$^\circ$ is required to enclose sufficient calibrator flux for the majority of nearby late M dwarf targets. The native $0.19^\circ$ field of the 1.3-m telescope at $f/$13.5, assuming a $4096 \times 4096$ detector with 15~$\mu$m pixels, falls short of this requirement. We designed and fabricated a four-lens focal reducer that increases the field to $0.48^\circ$ at $f/$5.6, providing a plate scale of 0$\farcs$43~pixel$^{-1}$.

The optical design has four spherical lenses fabricated from BSM51Y, CaF$_2$, S-TIH6, and S-BAH27 glasses, selected to minimize chromatic aberration while achieving high throughput in our narrow bandpass. A small lens-to-detector separation is highly favored, driving key architectural decisions: the fourth lens serves as the CCD dewar window and the detector operates in frame transfer mode to avoid the need for a inconvenient mechanical shutter. Custom anti-reflection coatings optimized for our 834--893~nm bandpass achieve $R_{\mathrm{max}} < 0.75\%$ across seven of the eight lens surfaces, with the exception of the concave back surface of lens 3 where the pronounced radius of curvature drives $R_{\mathrm{max}}$ up to 2.5\% at the steepest angles of incidence ($>45^{\circ}$); we refer the reader to \citetalias{GaMe23} for the full coating characterization.

\begin{figure*}[ht]
    \centering
    \includegraphics[width=\textwidth]{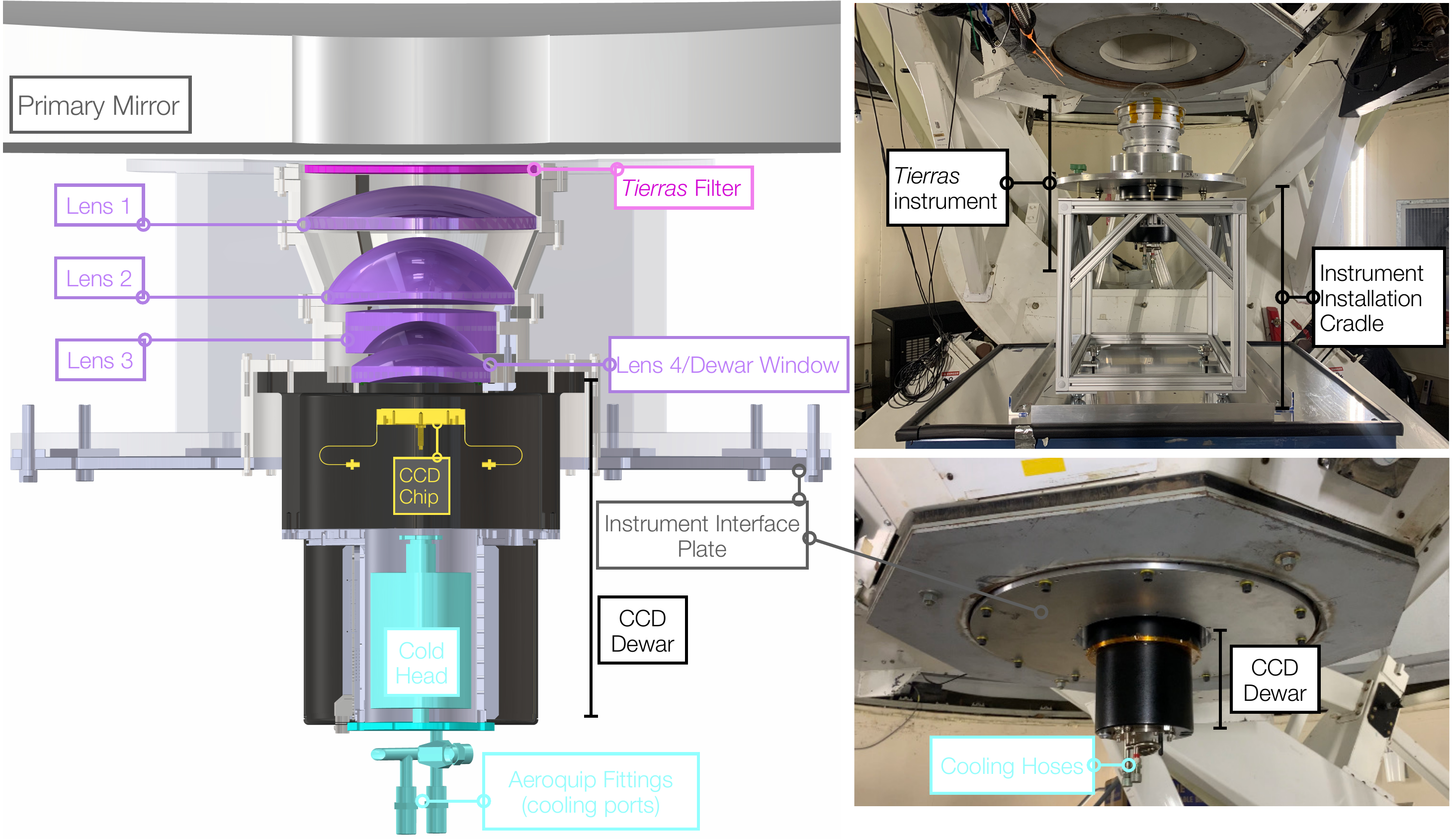}
    \caption{Graphical overview of the \textit{Tierras} instrument and its installation. \textit{Left:} CAD cross-section of the full \Tierras instrument installed on the 1.3-m telescope. From top to bottom, the telescope's native $f/13.5$ beam (not shown) crosses the primary mirror hole (white), passes through the \Tierras filter (pink) and lenses 1--4 (purple), producing an $f/5.6$ beam. The beam is imaged at the CCD chip (yellow), which is housed in a vacuum chamber inside the dewar (black). Components of the 1.3-m telescope are shown with reduced opacity, whereas opto-mechanical components of our instrument are shown in white. The instrument interface plate is shown in silver. \textit{Top right:} The focal reducer and field flattener fully integrated with the CCD dewar, resting atop our custom installation cradle and positioned underneath the primary mirror of the telescope (not visible in this view), immediately prior to installation. \textit{Bottom right:} The \Tierras instrument following installation on the 1.3-m telescope. Only the bottom of the CCD dewar is visible from this vantage point, as the rest of the instrument is nested immediately underneath the primary mirror.}
    \label{fig:optics_installed}
\end{figure*}

Each lens is bonded to a custom aluminum bezel using Momentive RTV560 silicone rubber, which retains its elastometric properties over the $-10^\circ$C to $+25^\circ$C operating temperature range. The bezels are located with precision dowel pins, enabling disassembly and reassembly with repeatable centration to within 6~$\mu$m (useful for cleaning the optics and carrying out upgrades periodically, \S \ref{sec:commissioning}). All lenses met the $\pm$25~$\mu$m decenter and axial displacement tolerances established during the design phase. We refer the reader to \citetalias{GaMe20} and \citetalias{GaMe23} for a complete description of the optical design, tolerance budget, fabrication, and assembly. The completed assembly is shown installed on the 1.3-m telescope in Figure~\ref{fig:optics_installed}.

\subsubsection{CCD Camera Sub-Assembly}\label{subsec:ccd}

\begin{figure*}[ht]
    \centering
    \includegraphics[width=\textwidth]{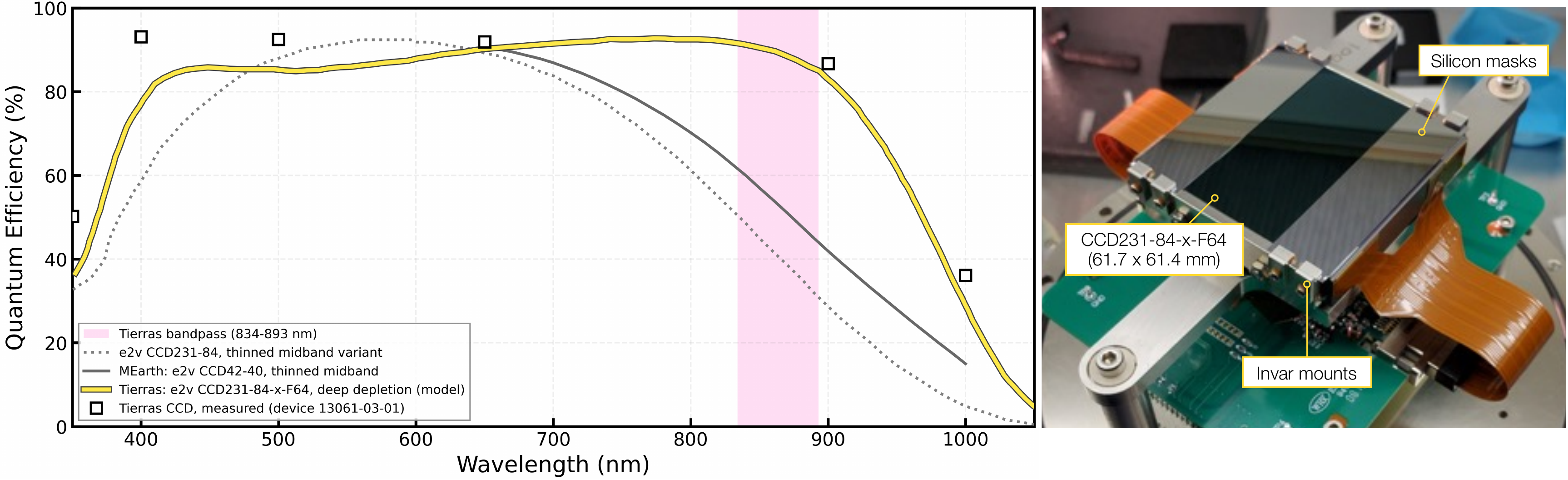}
    \caption{\textit{Left:} QE of the \Tierras deep-depletion CCD231-84 model curve provided by Teledyne (yellow), together with their acceptance measurements of our own device, serial 13061-03-01 (open squares). For comparison we show the back-illuminated, midband-coated
    e2v CCD42-40 used by MEarth (solid grey; \citealt{Nu08, Ir15}) and a thinned midband variant of the CCD231-84 itself (dotted grey). Averaged across the \Tierras bandpass (shaded pink), the deep-depletion device delivers 1.7 to 2.2 times the QE of the two comparison devices. \textit{Right:} The \Tierras CCD photographed during integration at
    Spectral Instruments in 2021 October, prior to installation in the dewar. The silicon masks that cover the two storage regions flank the central imaging area and are held via Invar mounts. They are aligned to within one pixel. The masks appear bright because the photograph predates their Aeroglaze Z306 coating; coated they are better than $99.999\%$ opaque.}
    \label{fig:ccd_mosaic}
\end{figure*}

\Tierras uses a Teledyne e2v CCD231-84 deep-depletion device with 4112$\times$4096 pixels (15~$\mu$m pitch). The deep-depletion architecture provides quantum efficiency (QE) exceeding 85\% in the \Tierras bandpass, roughly 1.7$\times$ higher than the back-illuminated, midband-coated devices used by comparable facilities such as MEarth \citep{Nu08, Ir15}. Teledyne's acceptance measurements of our device return QE values of 91.9\% at 650~nm and 86.7\% at 900~nm, the two test wavelengths bracketing the \Tierras bandpass (Figure~\ref{fig:ccd_mosaic}). Our pixel full well is 385,000~e$^-$ (Teledyne e2v, blooming limit, measured at device level), but the 16-bit analog to digital converter saturates first, at 65,535~ADU, so in practice the pixels are never fully filled. The device was classified as Grade~5 by the manufacturer owing to its 2,583 point defects and its dark current rather than by its photometric response: the device meets the Grade~0 specification for QE at every test wavelength, for charge transfer efficiency on all three functioning outputs, and for pixel response non-uniformity. Neither driver of the grade affects the measurements presented here. The defective pixels and columns are masked during reduction (Section~\ref{subsec:sci_reduction}), and the dark current contributes less than 0.03\% of the total noise at our typical exposure times (Section~\ref{subsec:sci_reduction}).

We operate the CCD in frame transfer mode, where accumulated charge is rapidly transferred into masked storage regions (covered by 200~$\mu$m-thick silicon wafers coated with Aeroglaze Z306) in 57~ms, after which the storage area is read out through two amplifiers at 1.39~MHz (read noise 18.5~e$^-$ RMS). As we show in Figure~\ref{fig:ccd_mosaic}, our device is split, with masked storage on both sides of the imaging area. Each half of the image travels only 1024 rows rather than the full parallel extent of the detector,doubling readout speed relative to conventional frame transfer devices. At the manufacturer's stated parallel shift time of 55~$\mu$s per row this corresponds to 56~ms. We emphasize that \Tierras does not currently overlap integration with readout: each exposure is commanded individually, so the array is cleared, integrated, transferred and read out in sequence, and the frame transfer acts as a fast electronic shutter rather than as a means of uninterrupted observation. The cadence is therefore the exposure time plus a fixed 7.0~s of transfer, readout and turnaround overhead, giving a per-exposure duty cycle of 81\% at our typical 30~s exposure and 90\% at 60~s. This is distinct from the night-level observing efficiency of 61\% we report in Section~\ref{subsec:obs_stats}, which also accounts for slews, acquisition sequences and focus runs.

The \Tierras camera subsystem was designed and built by Spectral Instruments (SI), with headquarters conveniently located in Tucson, AZ. The 25.6~mm airspace behind Lens~4 required a custom camera head, which we designed jointly. In the standard SI 1110 the CCD sits 34.45~mm behind a flat fused-silica window; to use Lens~4 as the window instead, the detector was moved 22.29~mm forward, to $12.160 \pm 0.025$~mm from the lens. The redesigned dewar front registers the Lens~4 bezel against a flat toleranced to 0.013~mm and carries a $1.91 \pm 0.03$~mm groove for the O-ring that seals the vacuum ($<10^{-4}$~psi against 10.9~psi ambient) directly against the pitch-polished back of the lens. A custom bolt pattern allows the dewar to be removed from the telescope without dismounting the optics.

The CCD operates at $-105^\circ$C, and is cooled by a closed-loop cryogenic system (also built by SI) which uses PT-30 refrigerant (a proprietary mix of noble gasses) that eliminates the need for liquid nitrogen refills and facilitates fully robotic operation.

Our pipeline adopts a gain of 5.9~e$^-$~ADU$^{-1}$ and a read noise of 18.5~e$^-$, derived together from a single calibration in 2021; the camera vendor's acceptance test and a photon transfer curve restricted to the low-signal regime both indicate a gain nearer 5.1~e$^-$~ADU$^{-1}$, so absolute conversions to electrons in this work may be high by ${\sim}15\%$. However, we note that the precisions we report are relative measurements from normalized light curves and are thus unaffected. 

\subsubsection{Telescope Baffles}\label{subsec:baffles}

The \textit{Tierras} 1.3-m telescope has an open-tube Cassegrain configuration in which direct skylight can bypass the mirrors entirely and reach the focal plane. To suppress this stray light, we designed and fabricated custom tubular baffles for both the primary (M1) and secondary (M2) mirrors following the graphical optimization of \citet{Pr68}, implemented via our sequential Zemax model and a custom Python routine 
(\href{https://github.com/jgarciamejia/tierras_baffles/blob/main/Code/baffle_optim.py}{\texttt{baffle\_optim.py}}; publicly available on GitHub). We adopted a tolerance budget of $2.5$--$5$~mm, consistent with that used for the MMT $f/5$ baffles \citep{Ca08}. The resulting for-fabrication dimensions are summarized in Table~\ref{tab:tierras_instrument}; full design details, including the nine annular disk baffles installed inside the M1 baffle tube to suppress single-bounce scattering, are provided in \citetalias{GaMe23}. 

\begin{figure}
    \centering
    \includegraphics[width=.95\columnwidth]{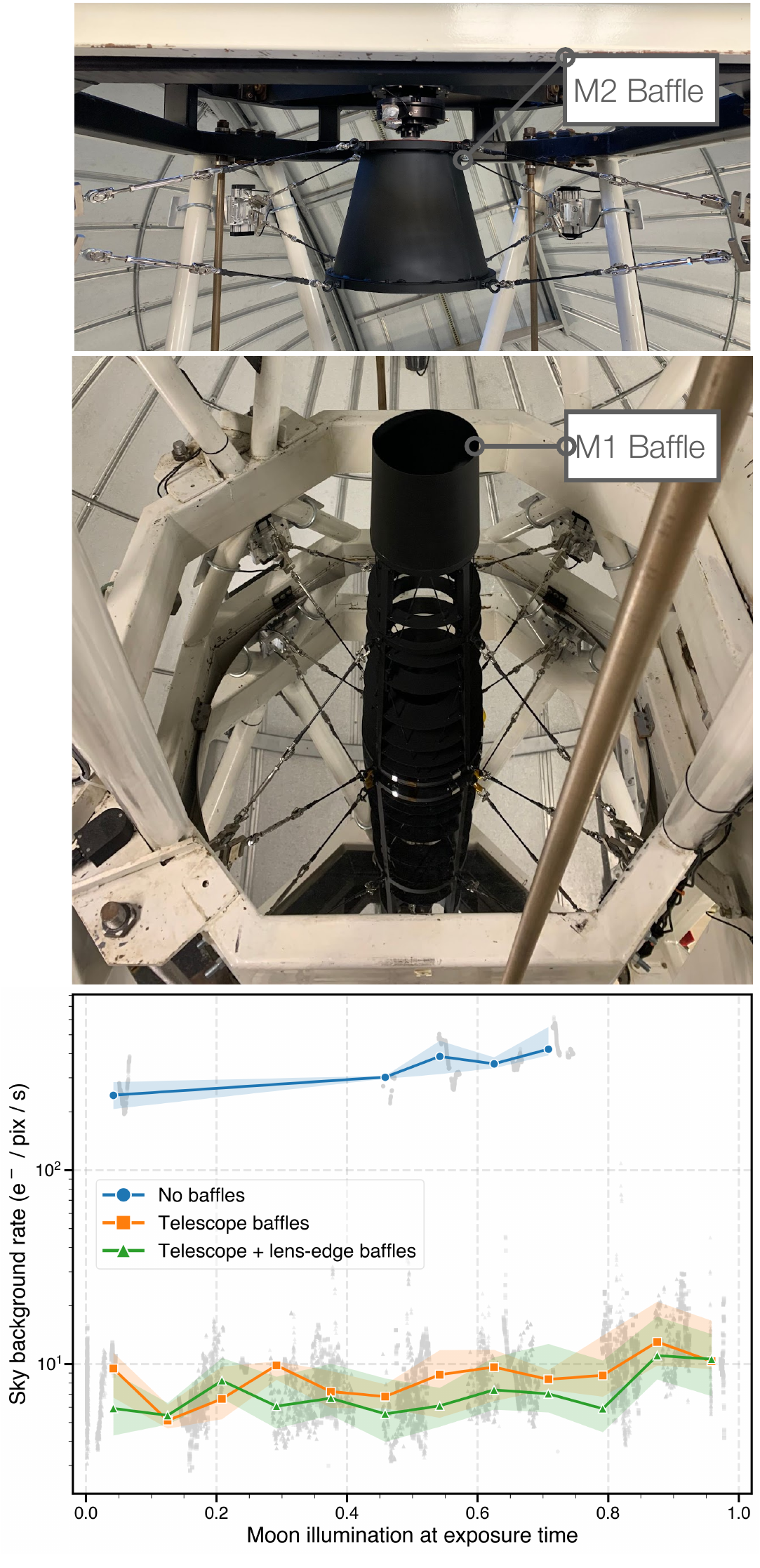}
    \caption{The \textit{Tierras} baffle system and its impact on sky background. \textit{Top panel:} the upper (M2) baffle pictured from its side. \textit{Middle panel:} the lower (M1) baffle with cable assemblies visible, and the primary mirror underneath it. \textit{Bottom panel:} median sky background rate as a function of lunar illumination for the three baffle configurations: no baffles (blue circles, 6 nights, 4303 exposures), telescope baffles only (orange squares, 18 nights, 6042 exposures), and telescope plus lens-edge baffles (green triangles, 19 nights, 8762 exposures). Light gray markers show individual exposures, while the bold curves show medians computed within 12 equal-width lunar-illumination bins from 0 to 1. We omit bins containing fewer than 10 exposures. The shaded bands indicate the 16th--84th percentile range of sky values within each bin.}
    \label{fig:baffles}
\end{figure}

Each baffle is suspended from the optical tube assembly via eight pre-tensioned wire rope cable assemblies, with one load cell per cable to enable real-time monitoring of cable tension (Figure~\ref{fig:baffles}). We initially coated the baffles in flat-black primer and paint; however, this proved highly reflective in the \textit{Tierras} NIR bandpass \citep{Schmidt2018,Schmidt2022}. We subsequently relined the baffle surfaces with IR Flock Sheet (Koyo Orient Japan Co., Ltd.\footnote{\url{https://koyo-orient.co.jp/en/}}), a flocked fabric engineered for high NIR light absorption. We installed the baffles on the telescope between 2022 February 15 and March 24.

To quantify the impact of the M1 and M2 baffles, we compared the sky background as a function of lunar illumination before and after their installation, drawing 4303 pre-baffle exposures (from 6 science nights spanning lunar illuminations 0.05--0.75) and 6042 post-baffle exposures (from 18 nights spanning 0.00--0.98), with a range of lunar altitudes sampled within each illumination bin. Our pre-baffle data coverage is limited: of the 13 nights with raw frames available from this era, only 6 contributed science exposures to this analysis. The remainder were commissioning observations or had unreliable astrometric solutions; we included only pre-baffle science observations with reliable astrometry. Because these nights predate the current photometry pipeline (Section~\ref{sec:pipeline}), we measured their median sky background per pixel from a sigma-clipped median of the entire image. For post-baffle nights we adopted a median of the pipeline's per-source local-annulus values; we verified that the two methods agree to within about $2\%$ on overlapping data. All sky values were converted to electrons using a detector gain of 5.9~e$^{-}$/ADU, and we restricted the analysis to frames taken at sun altitude below $-18^\circ$ to minimize twilight contamination.

As we show in the bottom panel of Figure~\ref{fig:baffles}, the M1 and M2 baffles reduce the median sky background by a factor of approximately 42, from 363 to 8.6~e$^{-}$~pixel$^{-1}$~s$^{-1}$. The baffles only somewhat flatten the facility's sky background dependence on lunar illumination, demonstrating that the pre-baffle sky was dominated by instrumental scatter rather than direct moonlight. We also measure a further approximately 21\% reduction at lunar illuminations $\geq 0.4$, realized after the lens-edge baffles were installed in 2025 August, which we describe in Section~\ref{subsec:inner_baffles}.

\begin{table*}[!t]
\centering
\caption{Key specifications for the four \Tierras instrument components. Full design details are provided in \citetalias{GaMe20} and \citetalias{GaMe23}.}\label{tab:tierras_instrument}
\footnotesize
\begin{minipage}[t]{0.47\textwidth}
\centering
\begin{tabular}{ll}
\hline
\hline
Parameter & Value \\
\hline
\multicolumn{2}{c}{\textit{Custom Bandpass Filter}} \\
\hline
Vendor & Asahi Spectra Co., Ltd. \\
Operating focal ratio & $f$/13.5 \\
Center wavelength ($\lambda_C$) & 863.5~nm \\
FWHM bandwidth & 40.2~nm \\
Peak transmission & $>$99\% \\
Out-of-band rejection & $<$0.1\% \\
& (300--834~nm, 893--1100~nm) \\
Transmission uniformity & $\pm$2\% across clear aperture \\
Substrate & Fused silica (OHARA SK-1300) \\
Substrate thickness & 6.15~mm \\
Mechanical diameter & 180.3~mm \\
Clear aperture & 174.4~mm \\
Design & 8 Fabry-P\'{e}rot cavities \\
\hline
\multicolumn{2}{c}{\textit{Focal Reducer \& Field Flattener}} \\
\hline
Configuration & 4 spherical lenses \\
Lens materials & BSM51Y (Lens 1), CaF$_2$ (L2), \\
& S-TIH6 (L3), S-BAH27 (L4) \\
System f-ratio & f/5.6 \\
Field of view & 0$\fdg$48 $\times$ 0$\fdg$24 \\
Plate scale & 0$\farcs$43~pixel$^{-1}$ \\
Lens 4--detector spacing & 25.6~mm \\
AR coating reflectivity & $<$0.75\% (834--893~nm) \\
\hline
\end{tabular}
\end{minipage}\hfill
\begin{minipage}[t]{0.47\textwidth}
\centering
\begin{tabular}{ll}
\hline
\hline
Parameter & Value \\
\hline
\multicolumn{2}{c}{\textit{Deep-Depletion CCD}} \\
\hline
Detector & Teledyne e2v CCD231-84 \\
Format & 4112$\times$4096 \\
& (2048$\times$4096 imaging) \\
Pixel size & 15~$\mu$m \\
Quantum efficiency & $>$85\% (in bandpass) \\
& (91.9,86.7\% at 650,900~nm) \\
Gain (pipeline value) & 5.9~e$^-$~ADU$^{-1}$ \\
Saturation (digitiser) & 65,535~ADU (16-bit) \\
Full well (blooming limit) & 385,000~e$^-$ \\
Read noise & 18.5~e$^-$ RMS (3.14~ADU) \\
Dark current & 0.00133~e$^-$~pixel$^{-1}$~s$^{-1}$ \\
& (vendor, $-105^\circ$C) \\
Charge transfer efficiency & $\geq$0.99999 \\
Pixel response non-uniformity & 1.5--1.7\% \\
Readout rate & 1.39~MHz \\
Frame transfer time & 57~ms (1024 rows) \\
Duty cycle & 81\% (30~s), 90\% (60~s) \\
Operating temperature & $-105^\circ$C (168.0~K) \\
%& within 1~K on 95\% of frames \\
Cooling system & Closed-loop cryogenic \\
& (PT-30 refrigerant) \\
\hline
\multicolumn{2}{c}{\textit{Baffles}} \\
\hline
M1 baffle height ($h_{1B}$)     & 1093~mm \\
M1 baffle top diameter ($D_{1B}$) & 200~mm \\  
M2 baffle height ($h_{2B}$)     & 315~mm \\
M2 baffle bottom diameter ($D_{2B}$) & 381~mm \\ 
Number of M1 annular disks      & 9 \\
Suspension                      & 16 pre-tensioned cables \\ 
& (8 per baffle) \\ 
Load monitoring                 & 16 load cells (1 per cable) \\    
Coating                         & IR Flock Sheet \\
& (Koyo Orient Japan) \\ 
Sky background reduction        & $\sim$42$\times$ (median) \\
\hline
\end{tabular}
\end{minipage}
\end{table*}

\subsection{Computing Infrastructure} \label{subsec:computers}

The \textit{Tierras} computing infrastructure consists of two computers housed in a 22U rack inside the control room adjacent to our dome. The need for two computers is a constraint we inherited from the DFM refurbishment (Section \ref{sec:site}): the TCSGalil telescope control system requires Windows, whereas the camera software development kit (SDK) and all our custom software run on Linux. Our main computer, which we refer to as the Observatory Control Server (OCS), is a 1U Supermicro A+ server (AMD EPYC 7262 processor, 64~GB ECC RAM) running Ubuntu Linux 18.04, with 15~TB of NVMe solid-state storage across two redundant drives. Our second computer is a Windows~10 PC running the DFM TCSGalil telescope control system installed during the 2020 refurbishment \citepalias{GaMe20}. The CCD camera interfaces to the OCS via a PCI Express (PCIe) optical fiber add-in card and SDK provided by Spectral Instruments (Section \ref{subsec:ccd}). For security, our OCS does not accept inbound network connections; instead, we access the facility remotely via SSH-tunneling into a local VNC server. 

\begin{figure*}
    \centering
    \includegraphics[width=\linewidth]{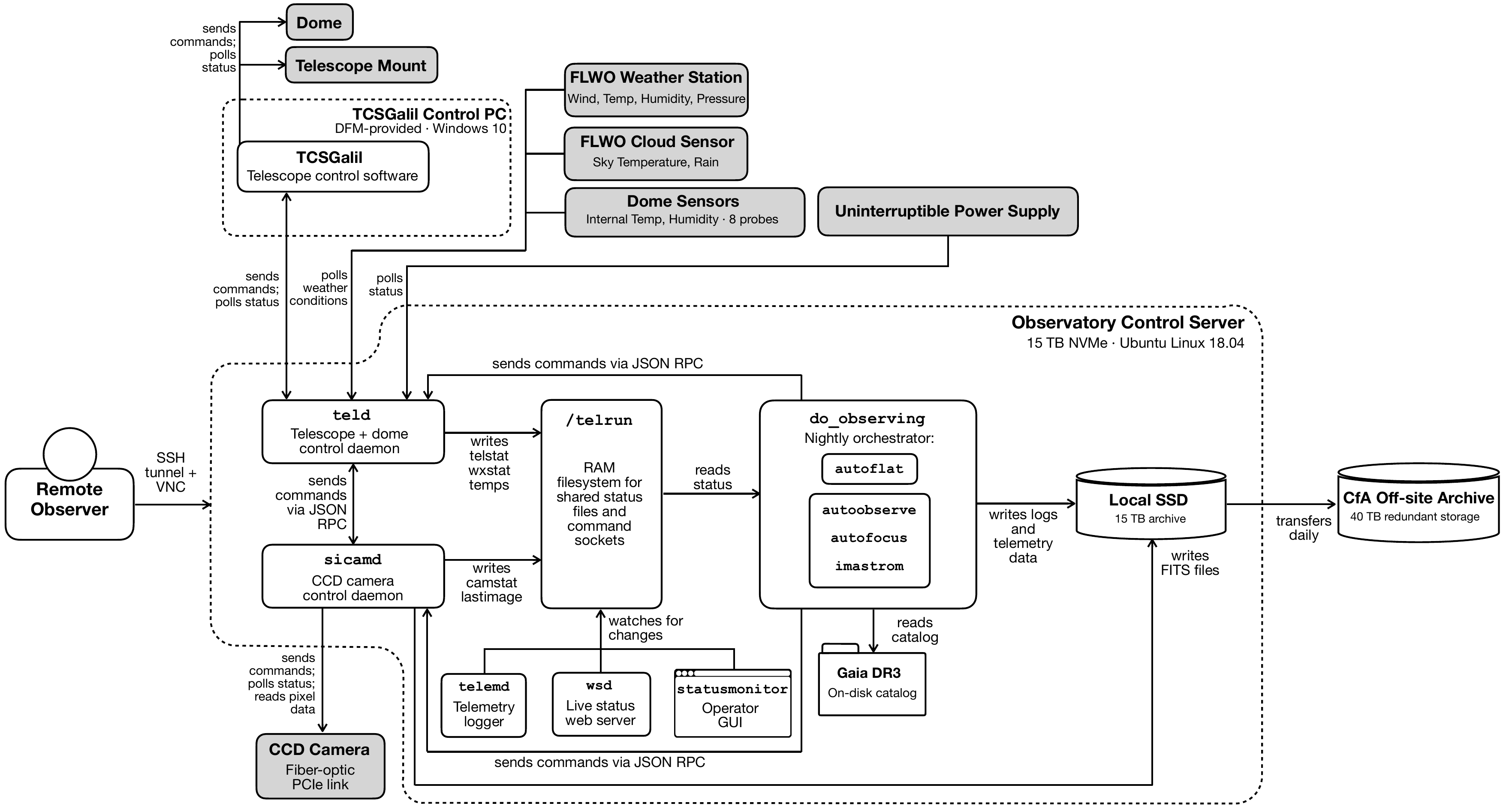}
    \caption{The \Tierras computing and control software architecture, laid out following the C4 model convention \citep{Brown2018_C4}. Our two main computers (the Observatory Control Server and the TCSGalil Control PC) are delimited by dashed boxes. Hardware sub-systems are shown in light gray, software processes in white, data stores as cylinders, and the on-disk catalog as a folder icon. The Observatory Control Server hosts two daemon processes called \texttt{teld} (telescope and dome control plus weather monitoring) and \texttt{sicamd} (camera control). Both processes publish status updates to \texttt{/telrun}, which is a shared RAM filesystem used for inter-process communication. Our nightly observations are orchestrated by the \texttt{do\_observing} script, which reads status from \texttt{/telrun} and sends commands to both daemons via JSON-RPC over Unix sockets. \texttt{do\_observing} coordinates nightly observations via its sub-modules \texttt{autoflat}, \texttt{autoobserve}, \texttt{autofocus}, and \texttt{imastrom}. Three auxiliary processes monitor \texttt{/telrun} for changes: \texttt{telemd} (telemetry logging), \texttt{wsd} (live web status server), and \texttt{statusmonitor} (operator GUI). Finally, \texttt{do\_observing} writes logs and telemetry data to the local SSD drives, whereas raw FITS images are written directly by \texttt{sicamd} to the local storage. A daily \texttt{cron} job is used to transfer data from our local storage to the CfA network storage.}
    \label{fig:architecture}
\end{figure*}

We operate \textit{Tierras} entirely remotely using control software modeled after the MEarth Observatory \citep{Ir15}. Our system architecture is diagrammed in Figure \ref{fig:architecture} and summarized below. Within the OCS, we control the facility through a set of communicating daemon processes. The camera daemon, \texttt{sicamd}, written in C using the Spectral Instruments SDK, receives commands via a JSON-RPC protocol over a UNIX domain socket and writes raw FITS images directly to the local solid-state drives (SSDs). A second daemon called \texttt{teld}, written in Python, translates the same JSON-RPC interface into TCP commands for the TCSGalil, to control and monitor both the mount and the dome. \texttt{teld} also monitors observatory safety: it polls weather conditions via HTTP from the FLWO site weather station, queries the uninterruptible power supply, and keeps track of the solar altitude, automatically closing the dome and stowing the telescope whenever any condition is unsafe. Both daemons publish real-time status snapshots (telescope state, camera state, and weather telemetry) to a shared RAM-based filesystem (\texttt{/telrun}). Three auxiliary processes (\texttt{telemd}, \texttt{wsd}, and \texttt{statusmonitor}) watch \texttt{/telrun} for changes to provide telemetry logging, live web status, and a graphical operator display, respectively. We utilize the shared RAM-based filesystem for two reasons: 1~Hz writes to our SSDs (i.e., the maximum rate at which we update status data) would wear them out, and RAM-based filesystems allow all processes to exchange status without conflicts.

Nightly operations are managed by a Python script called \texttt{do\_observing}. At the start of each night, the script opens the dome at evening twilight. If the sky is clear, as assessed by a measurement of the sky temperature, it then invokes \texttt{autoflat} to acquire sky flats by tracking the anti-solar point and setting exposure times to maintain an average flux level of $20,000$ ADU (\S \ref{subsec:flat_fields}). Once flats are complete, the script invokes \texttt{autoobserve}, our robotic scheduler and observer. Our current configuration of \texttt{autoobserve} divides the night into five-minute intervals and evaluates target observability subject to constraints on hour angle, zenith distance, and lunar proximity. For each interval, the scheduler selects either a target we have designated for high-priority staring via a \texttt{force\_obs.txt} configuration file, or the observable target from one of our science programs that has gone the longest without being observed. For each target, \texttt{autoobserve} slews the telescope and then performs a sequence of two acquisition exposures and telescope offsets to shift the target to the desired detector pixel. Each acquisition exposure is 10 seconds, after which  \texttt{imastrom} is invoked to plate-solve the image against a local copy of the Gaia~DR3 catalog\footnote{We maintain the catalog locally to avoid any network dependency during observations.}, solving for the offsets needed to position the field correctly. After the acquisition sequence, \texttt{autoobserve} then initiates the science exposure sequence, with guide corrections passed to the telescope after each image is read out.  At the beginning of observations and every three hours thereafter (as long as the outside temperature is above $0^{\circ}$ C, below which our secondary focus mechanism experiences significant stiction), \texttt{autoobserve} also invokes \texttt{autofocus} to perform a $V$-curve focus optimization. Between these calibrations, \texttt{teld} maintains focus continuously via a linear temperature–elevation compensation model. If weather or power conditions deteriorate at any point, \texttt{teld} closes the dome independently of the scheduler; \texttt{do\_observing} detects this, invokes \texttt{autoobserve} to re-plan the remainder of the night, and re-opens when and if conditions improve.

Whereas \texttt{do\_observing} writes observing logs and telemetry data to our local SSDs, \texttt{sicamd} is in charge of writing raw FITS images nightly into a directory tree on the local SSDs organized by the UT date. A daily \texttt{cron} job transfers all data to redundant 40~TB Network Appliance (NetApp) disk arrays managed by the Computation Facility at the CfA, where they are additionally backed up to an off-site facility. The photometric pipeline described in Section~\ref{sec:pipeline} operates on the NetApp archive.

\section{Commissioning \& Upgrades}\label{sec:commissioning}

\subsection{Installation and Collimation}\label{subsec:collimation}

We installed the \textit{Tierras} instrument on the 1.3-m telescope in 2021 July using a custom three-part lifting fixture comprising a 2000-lb-capacity hydraulic cart, a precision 80/20 aluminum cube for tip/tilt and centration adjustment, and a scissor-lift barrel for the dewar. The optical assembly is seated onto the telescope interface via 12 bolts. The total instrument mass is 86.5~lbs. First engineering light was obtained on 2021 July 28 with defocused images of M15 and NGC\,6940, which revealed significant coma indicating misalignment between the primary and secondary mirrors.

To characterize and correct the telescope collimation, we designed and built a custom laser-based metrology system consisting of a HeNe laser mounted on precision tip/tilt and $x$/$y$ linear stages, with centering jigs for the telescope instrument-interface surface, the primary mirror (M1), and the secondary mirror (M2). The system was calibrated in the laboratory to a sensitivity of 0.04$^\circ$ in tip/tilt and 0.5~mm in translation. During 2021 October, we used this system to align M1 and M2, correcting a 1.75~mm secondary decentration and a 0.1$^\circ$ tip/tilt offset. We also measured the residual tip/tilt between the instrument and M1 ($\sim$0.17$^\circ$ in both axes) and decentration (3.52~mm East, 1.42~mm South); Zemax modeling confirmed that these offsets degrade the RMS spot size by $\sim$15\% relative to perfect alignment, remaining below the typical seeing at Mt Hopkins (\S \ref{sec:seeing}). After collimation, defocused stellar images showed a centered secondary shadow with coma eliminated and field-dependent astigmatism greatly reduced. The images also showed six to seven zonal rings arising from polishing errors in the optics rather than from misalignment. We modeled these as a sinusoidal radial wavefront error using the LSST curvature wavefront sensing library \citep{Xi15} and recovered an amplitude of $\pm$15~nm. From the amplitude and spatial frequency we estimate the limiting optical FWHM of the system to be of order 0.4~arcsec, comparable to the 0.43~arcsec\,pixel$^{-1}$ plate scale. Since the typical seeing at the Ridge site is above 1~arcsec, the optics are not the limiting contribution to \Tierras image quality. The residual astigmatism is attributable to the primary mirror's radial support system, not to collimation error. Full details of the metrology system and collimation procedure are given in \citetalias{GaMe23}.

\subsection{Optics Upgrade: Inner Lens-Edge Baffles}\label{subsec:inner_baffles}

During the first three years of Tierras operations, all images exhibited a pervasive radial flux gradient at the focal plane, with a glow-to-sky flux ratio of 0.15--0.21 that was consistent across varying lunar illumination, source counts, and exposure times. This gradient prevented effective flat-fielding, and the photometric pipeline instead relied on local background subtraction using per-source annuli. In 2024 October, off-axis observations of Vega revealed glint artifacts that appeared at $\sim$0.30$^\circ$ off-axis and disappeared at $\sim$0.36$^\circ$, indicating that the scattered light originated from the chamfered edge of Lens~3 (L3) in the focal reducer/field flattener assembly. Zemax ray-trace modeling confirmed that rays at field angles of 0.302$^\circ$ and $-$0.357$^\circ$ intercept the L3 leading edge, scattering light onto the detector. A temporary 128-mm-aperture mask installed over the M1 central hole in 2024 November confirmed the diagnosis: the glow was entirely suppressed.

\begin{figure*}
    \includegraphics[width=\linewidth]{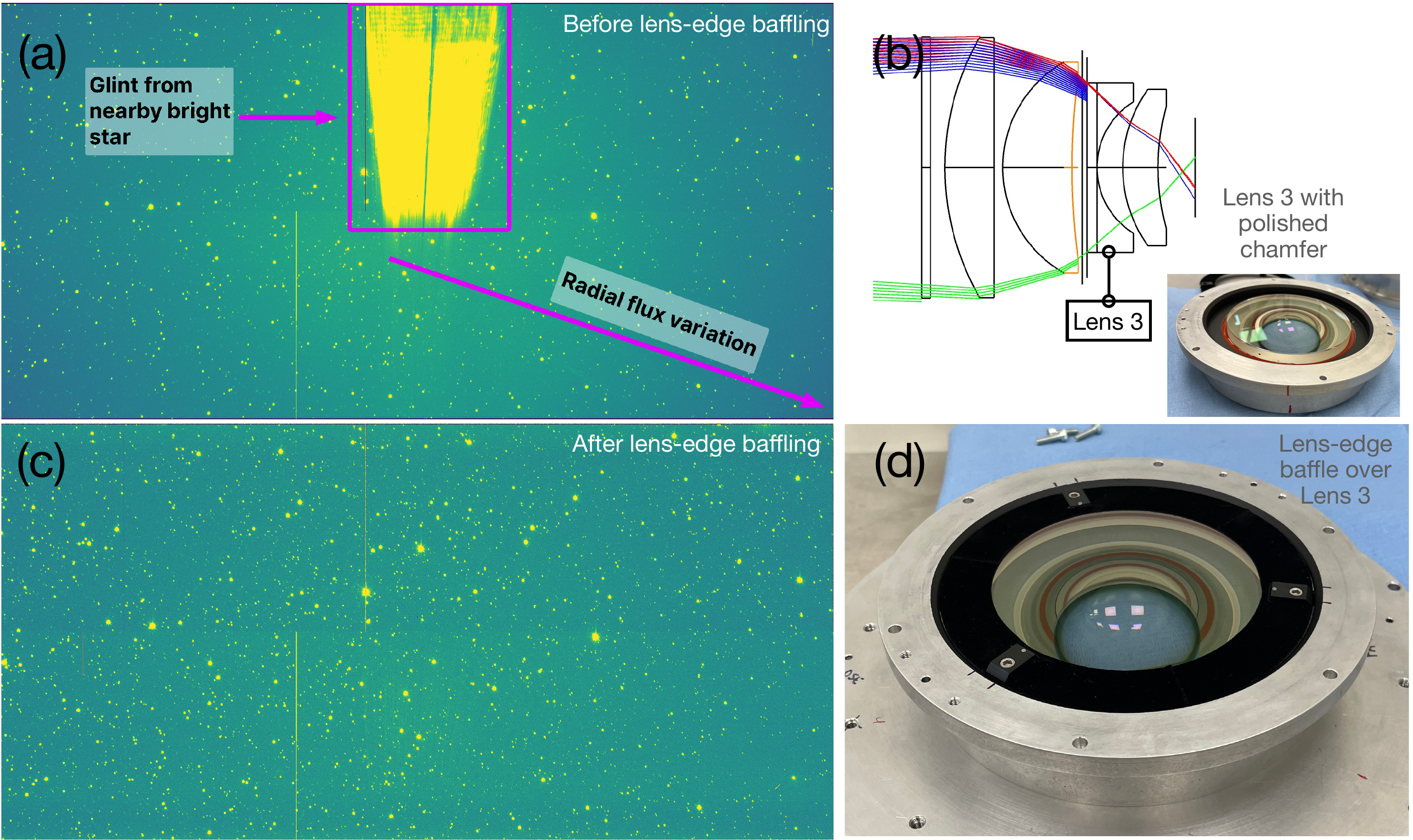}
\caption{Diagnosis and correction of intra-instrument scattered light affecting \textit{Tierras} images taken between commissioning and 2025 August. (a) A \textit{Tierras} image taken prior to the installation of lens-edge baffles interior to the focal reducer and field flattener. This field lies near a bright star ($\beta$~Cas, $V=2.3$~mag) and exhibits two compounding scattered light artifacts: a radial decrease in background flux away from the image center (present in \textit{all} \Tierras images prior to the optics upgrade), and a bright glint caused by proximity to the bright field star. (b) Ray trace of the \textit{Tierras} corrector optics illustrating the source of the scattered light. We trace rays at semi-field angles of $0.302^{\circ}$ (blue), $0.330^{\circ}$ (red), and $-0.357^{\circ}$ (green), all of which exceed \Tierras's maximum design semi-field angle of $u_\mathrm{pr} = 0.28^{\circ}$ (\S\ref{subsec:baffles}). These rays cut on and off at the leading edge of Lens~3, where we model the surface as a $45^{\circ}$ chamfer (odd asphere) to approximate the as-built edge geometry, which we found to be rounded and polished rather than a sharp bevel. Inset: the chamfer of Lens~3 as inspected during the optical upgrade. (c) The same field as panel~(a) following installation of the lens-edge baffles. Both the radial scattered light gradient and the glint artifact are effectively removed. (d) A lens-edge baffle coated in IR flocking installed above Lens~3, pictured during the 2025 upgrade.}
\label{fig:inner_baffles}
\end{figure*}

We implemented a permanent fix during 2025 August. We removed the instrument from the 1.3-m telescope, detached the optical train from the CCD dewar, and shipped the optics to our laboratory in Cambridge, MA.  We disassembled the optical train and installed light-absorbing annuli on the edges of all four lenses. Choosing the annulus flocking material required some effort. The Tierras bandpass (834--893~nm) sits in a region of the spectrum where the typical visible-optimized `blackest' coatings in the market (e.g., Musou Black) lose performance. Materials engineered specifically for strong absorption in this regime are less commonly characterized in the astronomical literature. Guided by the reflectance catalogs of \citet{Schmidt2018, Schmidt2022}, we settled on IR Flock Sheet (Koyo Orient Japan Co., Ltd.), an electrostatically-flocked rayon fabric whose pile fibers are carbon-loaded to maintain strong absorption through the near-infrared. We installed annuli directly on the edges of Lenses~1--3, and mounted the Lens~4 baffle on the back of the Lens~3 bezel since L4 serves as the dewar window and remained on site. The annulus inner diameters came from the Zemax-modeled beam footprint at each lens surface, plus a 2~mm clearance margin: 170~mm for L1, 139~mm for L2, 109~mm for L3, and 85~mm for L4. Post-upgrade images show the glint artifacts are entirely suppressed, and the large-scale flux gradient drops from $\sim$10\% peak-to-peak to $\sim$2.5\% (Figure~\ref{fig:inner_baffles}). Median-combined sky flats now produce visibly flat images across the detector. This upgrade is distinct from the OTA baffles described in Section~\ref{subsec:baffles}, which address stray light external to the corrector assembly.

\section{Photometric Pipeline}
\label{sec:pipeline} 

The \textit{Tierras} photometric pipeline is a set of fully autonomous programs that runs once per day following data transfer from the telescope to a NetApp cloud storage space managed by the CfA. Here, we detail the various stages of the pipeline that convert raw images into light curves. 

\subsection{Flat Field Construction}
\label{subsec:flat_fields}
Sky flats are performed at evening twilight on every clear observing night, as assessed using a sky temperature threshold. During this sequence, the telescope is pointed toward the ``null point" as defined in  \cite{Chromey1996} where the sky illumination gradient across the FOV is minimized. The detector is then exposed with a 1-s exposure time every 15 seconds until the average counts are below $20,000$~ADU. At this point, the flat sequence begins, and we lengthen the exposure time between each exposure to maintain an average of $20,000$ counts in the images. We dither the telescope around the null spot by integer multiples of $\pm$30\arcsec~in both R.A. and Decl. each exposure  (i.e., offsets of $0\arcsec$, $+30\arcsec$, $-30\arcsec$, $+60\arcsec$, $-60\arcsec$, etc.) so that stars can be removed by subsequent median combination of the flats. The flats are obtained until the exposure time exceeds 30~s, typically resulting in 40\textendash 50 flats.

After data transfer completes, the first step of the \textit{Tierras} pipeline is to median combine the flats (if any were taken). Every raw \textit{Tierras} image comprises two $1049\times4196$-pixel sub-images, representing the two detector halves read out by two different amplifiers. The pixels that represent the illuminated region of each sub-image are [1:1024,~51:4146], resulting in a $1024\times4096$ image after trimming. We use the [1:1024,~4149:4196] detector overscan region (a $1024\times48$-pixel image) to evaluate the bias level as a function of row number. We take the median across the 48 columns of the bias image and fit a degree-8 polynomial to the resulting 1024-element array. This fit is then subtracted off each row of the detector image, and then the two detector halves are stitched together. 

We found it necessary to correct for the ``shuttering"\footnote{We use ``shuttering" in quotation marks because we operate the \textit{Tierras} detector in frame-transfer mode, as opposed to a mechanical shutter.} effect when normalizing the bias-subtracted flats, where the effective exposure time of each row increases opposite the readout direction as a result of a finite readout rate (i.e., rows near the center of the chip receive a slightly longer exposure time). We found that a model of the form

\begin{equation}
s_i = \begin{dcases}
    1 + \frac{i}{\nu t} & \text{if } 0\leq i < 1024 \\
    1 + \frac{2047-i}{\nu t} & \text{if } 1024 \leq i < 2048
\end{dcases}
\label{eq:shutter}
\end{equation}

\noindent was appropriate to account for this effect. Here, $s_i$ is the shutter model value for row $i$, which ranges from 0 to 2047, $t$ is the exposure time in seconds, and $\nu$ is an empirical rate constant, fitted to our flats, of $7\times10^4$~Hz. Each bias-subtracted flat is divided by the shutter model and then median normalized. The set of 40\textendash 50 flats is then median combined. 

If a new combined flat was generated, we then create a new ``super flat", which uses all the combined flats created within the last 60 days and which is actually used to flat field the data. We motivate this step with the following back-of-the-envelope calculation. Each pixel in an individual sky flat contains approximately $118,000$ e$^{-}$ ($20,000~\text{ADU} \times 5.9~\text{e}^{-}/\text{ADU}$), with an associated fractional photon noise of $3$~ppt ($1 / \sqrt{118,000}$). By combining 45 flats on a typical night, we reduce the fractional uncertainty on the individual pixel flux measurements to about 0.4~ppt ($1 / \sqrt{45\times 118,000}$). This is larger than the night-to-night precision limit of 0.25~ppt that \textit{Tierras} is designed to reach, and is thus unacceptably high. If we assume that the detector pixel sensitivities are uniform on the timescales of a few months (which we have verified is valid), we can combine flats across multiple nights to achieve an acceptable photon noise level. On average, we find that conditions are suitable for taking twilight sky flats on about 50\% of nights. Thus, combining 30 expected flats over a 60-day window, the fractional pixel flux measurement uncertainties will be 0.08~ppt ($1 / \sqrt{30\times 45\times 118,000}$), much lower than our theoretical noise floor. 

\subsection{Science Image Reduction}\label{subsec:sci_reduction}
Once any new combined flats and super flats have been created, the pipeline then reduces any new science images. Individual detector halves are trimmed, bias-corrected, and stitched together following the same procedure as the flats as described in Section~\ref{subsec:flat_fields}. Pixels and columns flagged in a static bad-pixel mask, which covers the column defects reported by the manufacturer together with additional columns we identified during commissioning, are excluded from the photometry. The detector halves are then divided by the most recent super flat and saved to disk. We do not correct for dark current. The camera vendor's acceptance test of this unit measured 0.00133~e$^-$~pix.$^{-1}$~s$^{-1}$ at the $-105^\circ$C operating temperature. An SI reanalysis of a 2021 pre-delivery dark sequence yielded a per-amplifier dark current of 0.004 to 0.008~e$^-$~pix.$^{-1}$~s$^{-1}$; this figure includes the detector's hot-pixel population and is as such an upper bound on the rate that survives our bad-pixel mask. We also do not currently correct for nonlinearity. Teledyne e2v measured a maximum departure of 0.2 to 0.3\% between 6,000 and 47,500~ADU on each of the three functioning outputs of this device, SI verified linearity to within 1\% between 9,300 and 57,700~ADU, and our own reanalysis of the 2021 laboratory flats detected no departure between 40 and 60,000~ADU at a systematic floor of 0.5\%.Thus, we purposefully set exposure times so that the peak counts from our targets are within the range where the detector is known to behave linearly to 0.5\% ($100-60,000$ ADU) but note that nonlinearity effects may be important for especially bright or especially faint stars in the fields.

\subsection{Source Selection and FWHM Measurement}\label{sec:seeing}
The pipeline then identifies the sources in each field for photometric analysis. We rely on the World Coordinate System (WCS) solutions of the images, generated at the telescope after image readout, and Gaia DR3 data to make this determination. We search for sources in a local copy of Gaia DR3 over the \textit{Tierras} FOV centered on each field pointing. We use the DR3 space motions to propagate the positions of these sources to the \textit{Tierras} epoch. We retain all sources with a Gaia $G_\text{RP}$ magnitude less than 17, for which we achieve a photometric precision of about 2.5\% in 60-s exposures. We cross-match these sources with a local copy of the  \cite{BailerJones2021} catalog to get their distances. 

\begin{figure}
    \centering
    \includegraphics[width=\columnwidth]{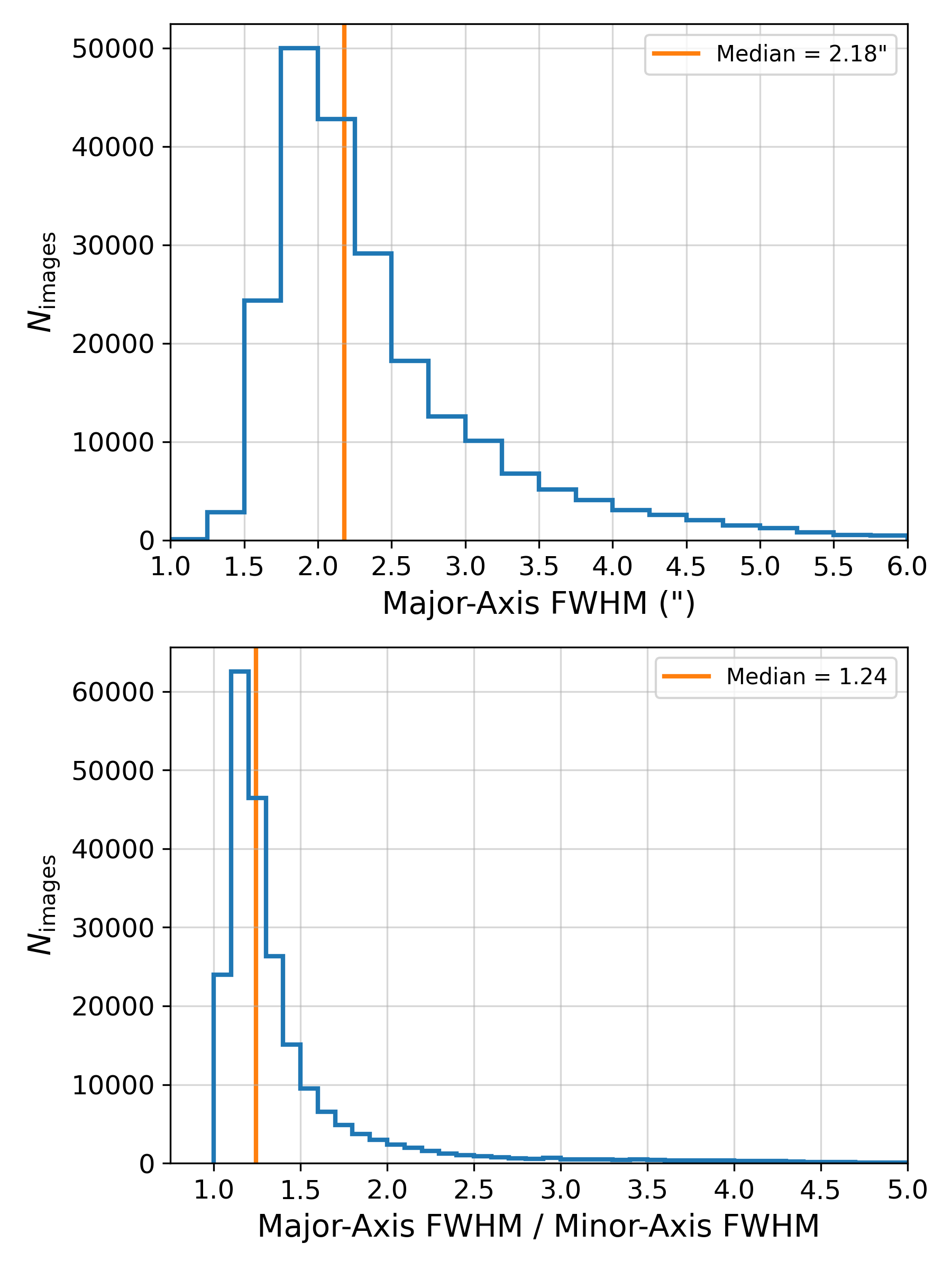}
    \caption{Top: The distribution of major-axis seeing FWHM values for images taken since 2023. The median seeing is 2.2\arcsec. Bottom: The distribution of the ratio between the major and minor FWHM axes. The median ratio is 1.2, and is attributable to tracking errors.}
    \label{fig:seeing}
\end{figure}

We then measure the FWHM seeing in each image. For each field, we identify the brightest star with $G_\text{RP} > 11$ (to avoid saturation) within $512\times512$-pixel boxes across the detector. We then fit 2D Gaussians to the resulting set of at most 32 stars, allowing the major and minor axes to differ. We convert these axes from $\sigma$ in pixels to FWHM in arcseconds and then take their averages across the set of fitted stars. We show the distribution of observed major-axis FWHM values since 2023 in Figure~\ref{fig:seeing}. The median major-axis seeing FWHM is 2.2\arcsec, and the median ratio between the major and minor axes is 1.2 (attributable to tracking errors). We note the long rightward tail of both distributions, which is mainly due to images with tracking failures, where sources become elongated in one direction.

\subsection{Photometry}
\label{subsec:photometry}
We use the WCS solution of each image to place circular photometric apertures on the source positions from Gaia DR3, propagated to the \textit{Tierras} epoch. We use apertures with radii ranging from 5\textendash 20 pixels in 1-pixel increments. These apertures were chosen such that their angular sizes (2.2\arcsec \textendash 8.6\arcsec) are approximately twice the typical FWHM seeing values experienced at the site (1\arcsec \textendash 4\arcsec; see Figure~\ref{fig:seeing}). 
We use the \texttt{aperture\_photometry} method from \texttt{photutils} \citep{photutils} to compute the sum of counts within each aperture (accounting for the exact overlap of the aperture on the pixel grid) and multiply by the detector gain $g=5.9~e^-\text{ADU}^{-1}$ to measure the raw source flux in electrons, $F_{\star,~ \text{raw}}$. 

We estimate the local sky background in each source aperture using circular annuli with an inner radius of 35~pixels and an outer radius of 55~pixels. We take the sigma-clipped mean of the distribution of pixel flux values in the annulus as the representation of the per-pixel sky background $F_\text{sky}$ for each source (after multiplying by the gain $g$), with a clipping level of 2$\sigma$. The background-subtracted flux $F_\star$ in the aperture of radius $R$ is then given by $F_\star = F_{\star,~ \text{raw}} - n_{\text{pix}}F_\text{sky}$ where $n_{\text{pix}} = \pi R^2$. We calculate the expected uncertainty on $F_\star$, $\sigma_{F_\star}$, using the photon noise contribution from the source and sky background, dark current, and read noise: 

\begin{equation} 
    \label{eq:ccd}
   \sigma_{F_\star} = 
  \sqrt{ 
    \begin{aligned}
        F_\star &+ n_\mathrm{pix}(1+\frac{n_\mathrm{pix}}{n_\mathrm{b}})(F_\mathrm{sky} + \phi_\mathrm{dark}t + \sigma^2_{\text{RN}})  \\
        &+(F_\star \sigma_\text{scint})^2
    \end{aligned}
  }
\end{equation}
where 

\begin{equation}
    \phi_\mathrm{dark} = 0.00133~\mathrm{e}^- \mathrm{pix}^{-1}\mathrm{s}^{-1}, 
\end{equation}

\begin{equation}
    \sigma_{\text{RN}} = 18.5~\mathrm{e}^-\mathrm{pix}^{-1}~\mathrm{RMS},
\end{equation}

\noindent $n_\mathrm{pix}$ is the area of the aperture in pixels, $n_\mathrm{b}$ is the number of pixels in the background annulus, $t$ is the exposure time in seconds, and 

\begin{equation}\label{eq:scint}
        \sigma_\mathrm{scint} = 0.135D^{-\frac{2}{3}}\chi^{\frac{7}{4}}(2t)^{-\frac{1}{2}}\mathrm{exp}(-\frac{h}{h_0})\sqrt{1+\frac{1}{n_R}}.
\end{equation}

\noindent This final expression is an estimate for the scintillation noise in relative photometry, where $D$ is the telescope diameter in cm (130~cm), $\chi$ is airmass, $t$ is the exposure time in seconds, $h$ is the elevation of the observatory (2306~m), $h_0=8000$~m, and $n_R$ is the number of uncorrelated reference stars, defined to be the number of references separated by more than $20\arcsec$ from the target \citep{Young1967, Dravins1998, Kornilov2012, Osborn2015, St17}. Because it is an estimate in relative photometry, it must be multiplied by the source flux to be appropriately scaled to units of photoelectrons. We emphasize that Equation~\ref{eq:scint} is an estimate, and that differences between measured and observed uncertainties for bright stars can likely be attributed to differences between this prescription and reality during the actual observations. We have ignored digitization noise in this expression which is negligible.

\subsection{Reference Star Weighting}

Reference stars are used in differential photometry to correct for variations that are common to the target in question and sources in the field. By dividing the measured target flux by the flux from some set of reference stars, the corrected flux will show variations from the target alone. However, not every star should be considered as a potential reference, since a reference star may itself be variable, or its photometry may suffer from systematics that are unknown beforehand. 

To account for these possibilities, we use a modified version of the reference star weighting scheme presented in \citet{Tamburo2022a} to create an ``artificial light curve" (ALC) to produce differential photometry for each source. This procedure iteratively determines the weights that should be assigned to each reference star in the ALC based on its own corrected light curve \citep{Broeg2005, Murray2020}.

We begin by assigning the $i^\text{th}$ star in the set an initial weight equal to the inverse of the median of its predicted variance over all frames (Eq.~\ref{eq:ccd}):
\begin{equation}
    W_{i} = \frac{1}{\mathrm{med.}(\sigma_{F_{\star,i}}^2)}.
\end{equation}
We repeat for all stars in the set and normalize the full array of weights so that they sum to one. 

We then initiate a ``crude" weighting loop which terminates when the differences between all weights in the current and previous loop are less than $10^{-4}$, or after 20 iterations. For the $i^\text{th}$ star, we do the following: 

\begin{enumerate}
    \item We set the weight of the star in question, $W_{i}$, to zero, since a star cannot be used as a reference for itself. We do the same for stars that have any frames in which the given aperture contains pixels above 40,000~ADU as a precaution against non-linear effects (Section~\ref{subsec:sci_reduction}). We then renormalize the weights array to sum to one. 
    \item We calculate the ALC in the $k^\text{th}$ frame for the $i^\text{th}$ star, given by 
    \begin{equation}\label{eq:alc}
        \mathrm{ALC}_{i,k} = \sum_{j=1}^{N} W_{j} F_{\star,j,k}
    \end{equation}
    where $j$ runs over the $N$ potential reference stars and $k$ indexes the frames. The ALC depends on $i$ because $W_i$ has been set to zero and the weights renormalized, so each star is corrected by a light curve built from all the others.
    
    \item We then calculate the $i^\text{th}$ star's relative flux in the $k^\text{th}$ frame using
    \begin{equation}\label{eq:rel}
        F_{\mathrm{rel},i,k} = \frac{F_{\star,i,k}}{\mathrm{ALC}_{i,k}}
    \end{equation}
    We median normalize the relative flux (that is, we divide the time series by its own median, so that the scatter measured in the next step is a fractional quantity comparable between stars) and perform a 4-$\sigma$ clipping to remove any significant outliers due to, e.g., cosmic ray hits. 
    \item We measure $\sigma_{\mathrm{meas},i}$, the standard deviation of $F_{\mathrm{rel},i,k}$ over all frames $k$, and replace $W_i$ with
    \begin{equation}
        W_i = \frac{1}{\sigma_{\mathrm{meas},i}^2}.
    \end{equation}
    This step moves us away from weights based on theory (Eq.~\ref{eq:ccd}) to weights based on actual measurements of light curve scatter.
    \item We repeat from Step 1 for the next star and keep iterating until the convergence threshold is met or until 20 iterations have been performed.
\end{enumerate}

After the crude loop has finished, we check for stars whose corrected flux is much noisier than expected based on $\sigma_{F_\star,i}$, indicating that they are significantly variable. We calculate the ratio of the final $\sigma_{\mathrm{meas},i}$ to the median-normalized $\sigma_{F_{\star,i}}$ for each source. We perform a sigma clipping on the resulting ratios and only retain the stars whose ratios are better than the 1-$\sigma$ upper bound of the clipped distribution as potential reference stars. Stars whose ratios are higher than the 1-$\sigma$ upper bound have their weights set to zero, and then the array of weights is again renormalized to sum to one. 

Finally, we perform a ``fine" convergence loop to determine the final reference star weights. This loop is functionally the same as the crude loop, but now we require the weights to converge to better than $10^{-6}$ (or 20 iterations) and the ALCs are computed excluding the most variable reference stars. The entire procedure is performed for all of the different aperture sizes that we used (described in Section~\ref{subsec:photometry}). For computational efficiency, we perform the weighting on binned data with 10 points per bin. 

\subsection{Light Curve Creation}

Once the reference star weights have been generated for all possible aperture sizes, we then create light curves for each source with each aperture size. We compute the final corrected light curve and expected uncertainty for a given aperture size for each star by doing the following.

We first calculate the ALC with Equation~\ref{eq:alc}. We calculate the uncertainty on the ALC using 
    \begin{equation}
        \sigma_\mathrm{ALC} = \sqrt{\sum_{j=1}^N (W_{j}\sigma_{F_{\star j}})^2}
    \end{equation}
    where $\sigma_{F_{\star j}}$ is the array of uncertainties calculated for reference star $j$ in each frame using Equation~\ref{eq:ccd}. Note that this and subsequent equations are evaluated on a per-target and per-frame basis, and that as such we have dropped the $i$ and $k$ subscripts for readability.

We then calculate the array of relative fluxes $F_\text{rel}$ following Equation~\ref{eq:rel} and calculate the associated uncertainty with 
    \begin{equation}
        \label{eq:sigma_rel_1}
        \sigma_\mathrm{rel} =\sqrt{\left ( \frac{\sigma_{F_\star}}{\mathrm{ALC}}\right ) ^2+\left ( \frac{F_{\star} \sigma_{\mathrm{ALC}}}{\mathrm{ALC}^2} \right ) ^2 } 
    \end{equation}\\

Finally, we median-normalize $F_\mathrm{rel}$ and its associated uncertainty, dividing each by $\widetilde{F}_\mathrm{rel} \equiv \mathrm{med.}(F_\mathrm{rel})$. Because the ratio $F_\star/\mathrm{ALC}$ is roughly constant, we may approximate $\widetilde{F}_\mathrm{rel} \approx F_\star/\mathrm{ALC}$. It is then worth exploring how this normalization simplifies Equation~\ref{eq:sigma_rel_1}. Writing $\bar\sigma_{\mathrm{rel}}$ for the normalized uncertainty, we get

\begin{align}
    \bar{\sigma}_\mathrm{rel} &= \frac{\sqrt{(\frac{\sigma_{F_\star}}{\mathrm{ALC}})^2+(\frac{F_\star\sigma_\mathrm{ALC}}{\mathrm{ALC}^2})^2 }}{F_\star/\mathrm{ALC}}\\
    &= \sqrt{\left [ \left ( \frac{\sigma_{F_\star}}{\mathrm{ALC}} \right )^2+\left ( \frac{F_\star\sigma_\mathrm{ALC}}{\mathrm{ALC}^2} \right )^2 \right] \frac{\mathrm{ALC}^2}{F_\star^2} }\\
    &= \sqrt{\left ( \frac{\sigma_{F_\star}}{F_\star} \right )^2+\left ( \frac{\sigma_\mathrm{ALC}}{\mathrm{ALC}} \right )^2   } \label{eq:sigma_rel_norm}
\end{align}
    
Substituting the expression for $\sigma_{F_\star}$ from Equation~\ref{eq:ccd} into Equation~\ref{eq:sigma_rel_norm} gives

\begin{equation}
    \label{eq:relative_sigma}
  \bar{\sigma}_\mathrm{rel} = 
  \sqrt{
    \begin{aligned}
     \frac{1}{F_\star}&+\frac{n_\text{pix}\left(1+\frac{n_\text{pix}}{n_\text{b}}\right)\left(F_\text{sky}+\phi_\text{dark}t+\sigma_\text{RN}^2\right)}{F_\star^2}    \\
     & + \sigma_\text{scint}^2 + \left ( \frac{\sigma_\mathrm{ALC}}{\mathrm{ALC}} \right ) ^2
    \end{aligned}
  }
\end{equation}

We can now examine the limiting dependence of Equation~\ref{eq:relative_sigma} on $F_\star$. For faint sources ($F_\star \to 0$), it reduces to 

\begin{equation}
    \label{eq:faint_limit}
   \bar{\sigma}_\text{rel} = \sqrt{\frac{n_\text{pix}\left(1+\frac{n_\text{pix}}{n_\text{b}}\right)(F_\text{sky} + \phi_\text{dark}t + \sigma_\text{RN}^2)}{F_\star^2}}.
\end{equation}

\noindent As expected, the noise budget of faint stars is dominated by sky background and read noise, since dark-current is negligible at our typical exposure times (Section~\ref{subsec:sci_reduction}). Meanwhile, for bright sources ($F_\star \to \infty$), Equation~\ref{eq:relative_sigma} reduces to 

\begin{equation}
    \label{eq:bright_limit}
    \bar{\sigma}_\text{rel} = \sqrt{\left ( \frac{\sigma_\mathrm{ALC}}{\mathrm{ALC}}\right ) ^2 + \sigma_\text{scint}^2}. 
\end{equation}

\noindent Equation~\ref{eq:bright_limit} shows that the achievable relative photometric precision for bright stars does not continually improve, as a naive application of Equation~\ref{eq:ccd} would suggest. Instead, it approaches a noise floor that is defined by the scintillation noise and the ALC correction. The ALC term can be understood intuitively by considering that for very bright stars, the ALC will be constructed from far fewer counts than are available from the target star itself, and thus dividing $F_\star$ by the ALC in this case will inflate the noise of the target star to be commensurate with the ALC. We will explore this behavior in more detail in Section~\ref{subsec:intranight_precision}.

Once $F_\mathrm{rel}$ has been calculated for all aperture sizes, we choose the light curve that minimizes the measured scatter to be the ``best" representation of that source's flux over our data set. 
 
\section{On-Sky Performance} 
\label{sec:performance}

\subsection{Single-Night Photometric Performance}\label{subsec:intranight_precision}

Equation~\ref{eq:relative_sigma} gives the expected per-exposure noise of normalized photometry, with terms for the source photon noise, sky photon noise, dark current, read noise, ALC correction, and scintillation. We can compare this expectation with the actual measured noise performance of \textit{Tierras} light curves, quantified by the light curve standard deviation. We make this comparison using \textit{Tierras} observations of the Kepler-1658 field from UT~2025~June~07, during which we obtained a continuous set of exposures in clear conditions over approximately 4.5 hours at a 30-s exposure time.

For the purpose of visualization, we first calibrated the background-subtracted flux of sources as a function of their Gaia $G_\text{RP}$ magnitudes. We restricted this fit to sources with flux contamination estimates from nearby sources less than 0.1\% (see below) and iteratively fit a function of the form $c\cdot10^{-G_\text{RP}/2.5}\cdot t$ to the average fluxes, where the exposure time $t$ for these data  was 30~s. Each iteration, we performed a 4-sigma clipping on the measured fluxes divided by the fit and terminated the fitting once zero sources were removed. After the final iteration, our flux calibration had the form 

\begin{equation}
    \label{eq:flux_calibration}
     F_\star  \approx 9.7e8 \cdot 10^{-G_\text{RP}/2.5} \cdot t ~[\text{e}^-].
\end{equation}

\begin{figure}
    \centering
    \includegraphics[width=\columnwidth]{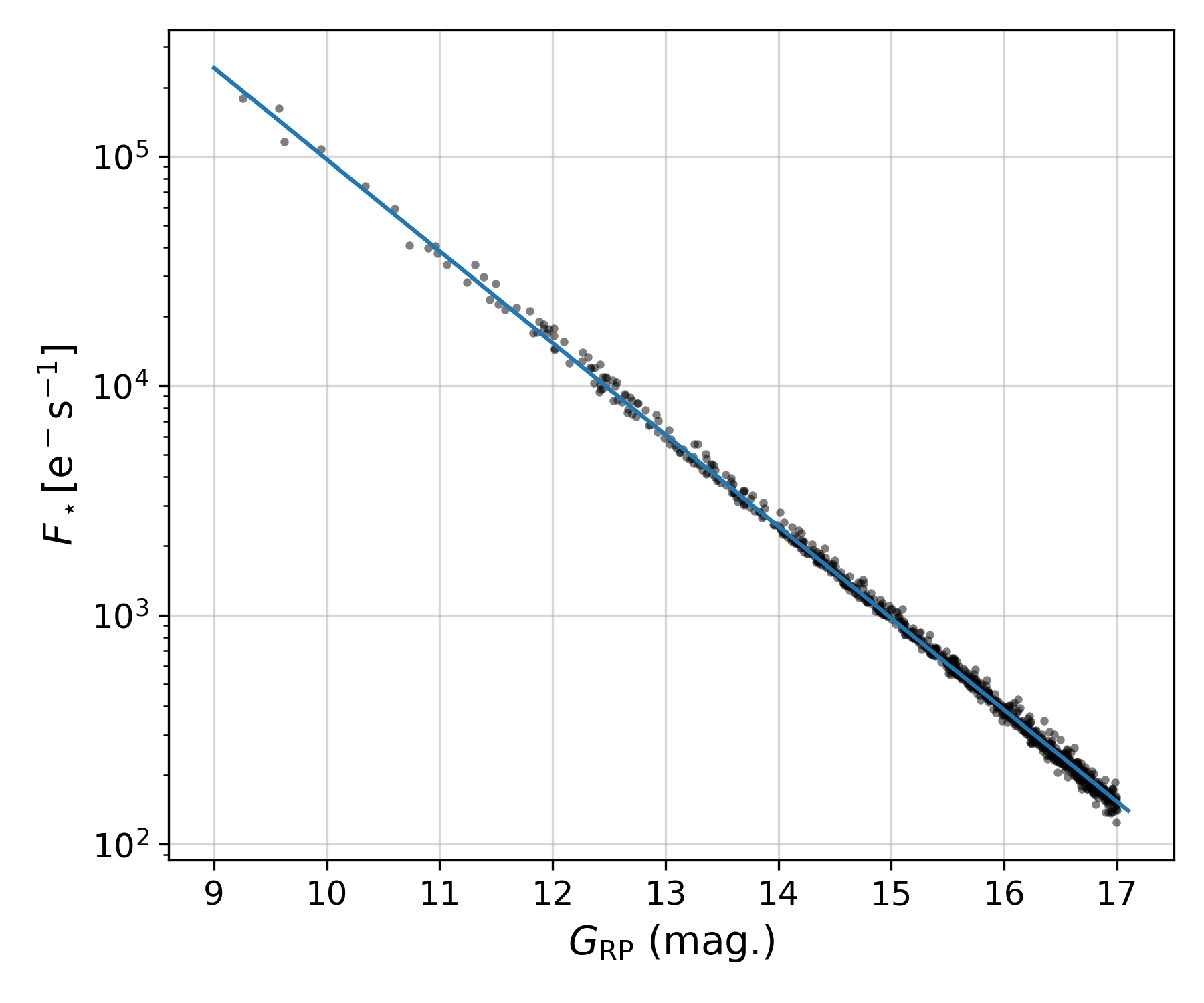}
    \caption{Calibration of average \textit{Tierras} fluxes for sources in the Kepler-1658 field versus their Gaia $G_\text{RP}$ magnitudes. The blue line shows a fitted function of the form $9.7e8\cdot10^{-G_\text{RP}/2.5}$ in units of e$^-$s$^{-1}$.}
    \label{fig:flux_calibration}
\end{figure}

We present a plot of the final flux sample and the fit in Figure~\ref{fig:flux_calibration}. This plot verifies that the slope of the measured fluxes in log-linear space is indeed $-2.5$, validating the substitution of Gaia $G_\text{RP}$ magnitudes as a stand-in for \textit{Tierras} fluxes. The expression for $F_\star$ from Equation~\ref{eq:flux_calibration} can then be substituted into Equation~\ref{eq:relative_sigma} to derive the expected average noise performance as a function of source $G_\text{RP}$ magnitude.

\begin{figure*}
    \centering
    \includegraphics[width=\textwidth]{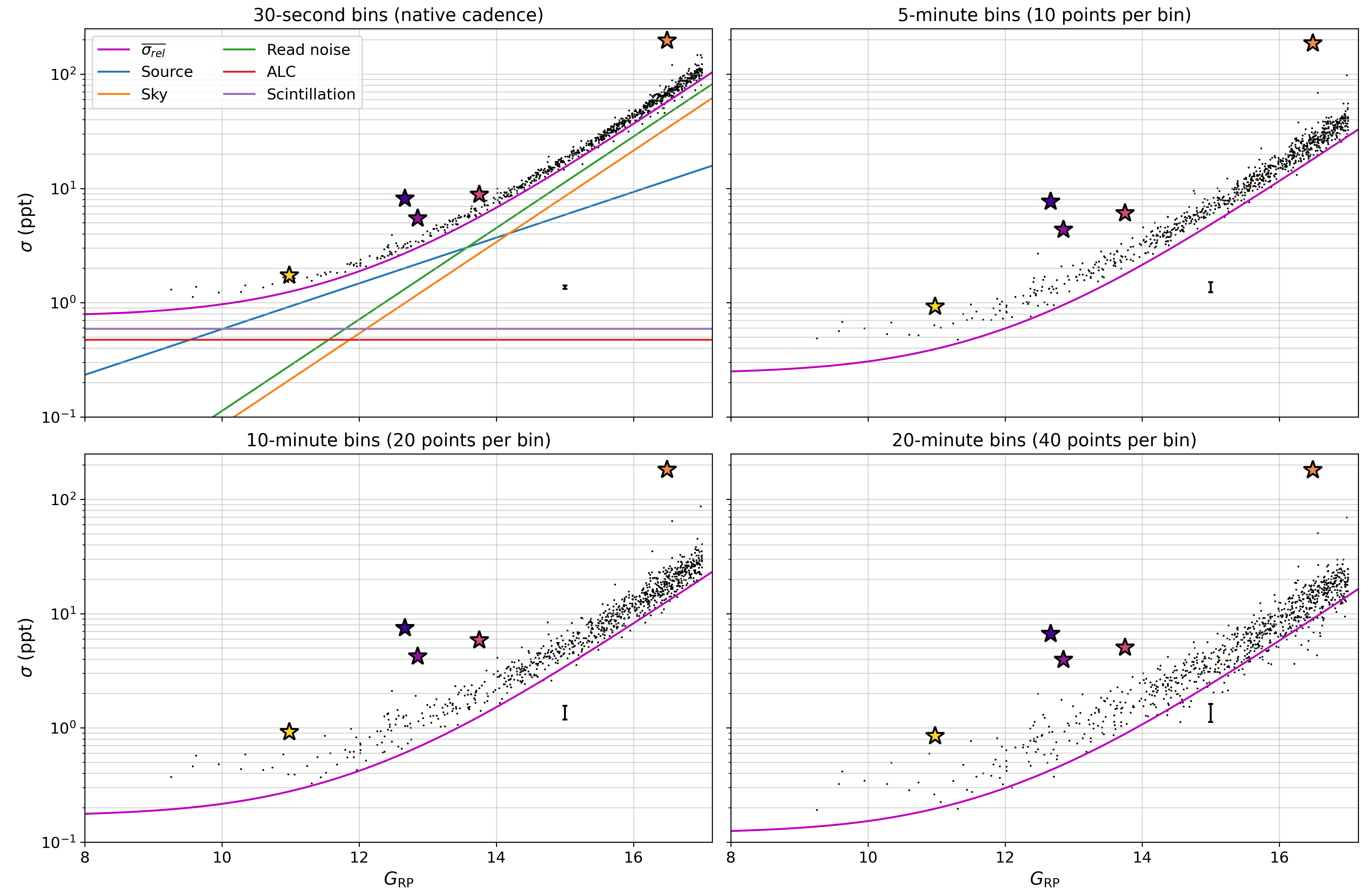}
    \caption{Standard deviation of sources in the Kepler-1658 field across a single night of continuous observations at different degrees of binning. Black points show measured light curve standard deviations in ppt, while the magenta line shows $\overline{\sigma}_\mathrm{rel}$ (Equation~\ref{eq:relative_sigma}). In the top-left panel, the individual components that contribute to $\overline{\sigma}_\mathrm{rel}$ are shown; they are omitted in the following panels for clarity. In each panel, an error bar in the lower right indicates the median fractional uncertainty on the light curve standard deviation across all sources as determined through a bootstrap simulation as described in the text. The stars indicate significantly variable sources whose light curves we show in Figure~\ref{fig:intranight_companion}. Top left: native cadence light curves. Top right: 5-minute bins. Bottom left: 10-minute bins. Bottom right: 20-minute bins.}
    \label{fig:intranight_precision}
\end{figure*}

In Figure~\ref{fig:intranight_precision}, we present the measured standard deviations of \textit{Tierras} light curves as a function of source $G_\text{RP}$ magnitude for observations of the Kepler-1658 field on UT~2025~June~07 at different levels of binning: the native 30-second cadence along with 5-, 10-, and 20-minute bins. All light curves were computed with a 10-pixel-radius circular aperture and were sigma-clipped with a threshold of $4\sigma$. We assessed the contamination of each source by modeling the flux contribution into the photometric apertures from nearby stars, using their Gaia $G$ magnitudes\footnote{We used $G$ magnitudes for the contamination estimate instead of the $G_\text{RP}$ magnitudes used elsewhere in this work because more sources have $G$ magnitude measurements than $G_\text{RP}$.} and positions. We examined a 100-pixel radius region around each source and modeled the source and any neighboring sources using circular 2D Gaussian PSFs with $\sigma=1.67$, corresponding to the median seeing FWHM $1\farcs7$ observed on this night. We retained sources for the plot if they were contaminated by less than 0.1\%. A total of 980 out of 1821 sources with $G_\text{RP} < 17$~mag were retained.

For each light curve at each degree of binning, we conducted a bootstrap simulation to estimate the uncertainty on the measured standard deviation.  We resampled each light curve with replacement $10,000$ times, computing the standard deviation of each resampling. The standard deviation of the distribution of $10,000$ sample standard deviations then gives an estimate of the uncertainty on the standard deviation measurement for each light curve at the given bin size. In each panel of Figure~\ref{fig:intranight_precision}, we show the median fractional uncertainty calculated across all sources at the given level of binning. Because the plot is on a log scale, the median fractional uncertainty is scale independent. Every point in Figure~\ref{fig:intranight_precision} has a value $\sigma$, associated uncertainty $\varsigma$, and fractional uncertainty $\epsilon=\varsigma/\sigma$. In log space, the uncertainty occupies a vertical range of $\log{\left ( \sigma + \varsigma\ \right)} - \log{\left ( \sigma - \varsigma\right ) } = \log{\frac{\sigma + \varsigma}{\sigma - \varsigma }} = \log{\frac{\sigma (1+\epsilon)}{\sigma(1-\epsilon)}} = \log{\frac{1+\epsilon}{1-\epsilon}}$, which is independent of the value of $\sigma$. As this calculation shows, the uncertainty on the light curve standard deviation explains both the observed scatter at a given level of binning and the increased scatter as larger bin sizes (and hence fewer points) are used to compute the standard deviation. 

The total noise model from Equation~\ref{eq:relative_sigma} is shown along with the constituent terms to visualize how different noise sources contribute to the overall budget across the continuum of source brightnesses. The sky background, ALC, and scintillation noise terms were visualized using the average value for each term across the night. For very bright sources, $\overline{\sigma}_\mathrm{rel}$ is dominated by scintillation and the ALC correction, as predicted in Equation~\ref{eq:bright_limit}. For faint sources, it is dominated by read noise and sky background, as predicted by Equation~\ref{eq:faint_limit} (ignoring dark current, which is negligible). We emphasize that no fitting has been done to align the $\overline{\sigma}_\mathrm{rel}$ line with the observed standard deviations, demonstrating the accuracy of our noise model. 

The separate panels of Figure~\ref{fig:intranight_precision} show the standard deviations of light curves assessed at the native 30-s cadence, along with 5-minute, 10-minute, and 20-minute bins. Assuming Gaussian uncorrelated noise, binning the light curves should provide a $\sqrt{N}$ improvement, where $N$ is the number of points in each bin. With a 30-s exposure time, we thus expect a $\sqrt{10}$ improvement in 5-minute bins, a $\sqrt{20}$ improvement in 10-minute bins, and a $\sqrt{40}$ improvement in 20-minute bins. These precision improvements are observed, but they are subject to increased scatter due to sampling noise as fewer and fewer points are used to calculate the standard deviation (see above). We note that there are 6 sources for which we observed light curve scatters better than 0.3~ppt in 20-minute bins. Assuming a full transit time of 1 hour, this precision is sufficient to detect Earth around a $0.31~R_\odot$~M3.5 dwarf or Ganymede around a $0.13~R_\odot$~M6.5 dwarf, both to a significance of $5\sigma$.

\begin{figure*}
    \centering
    \includegraphics[width=\textwidth]{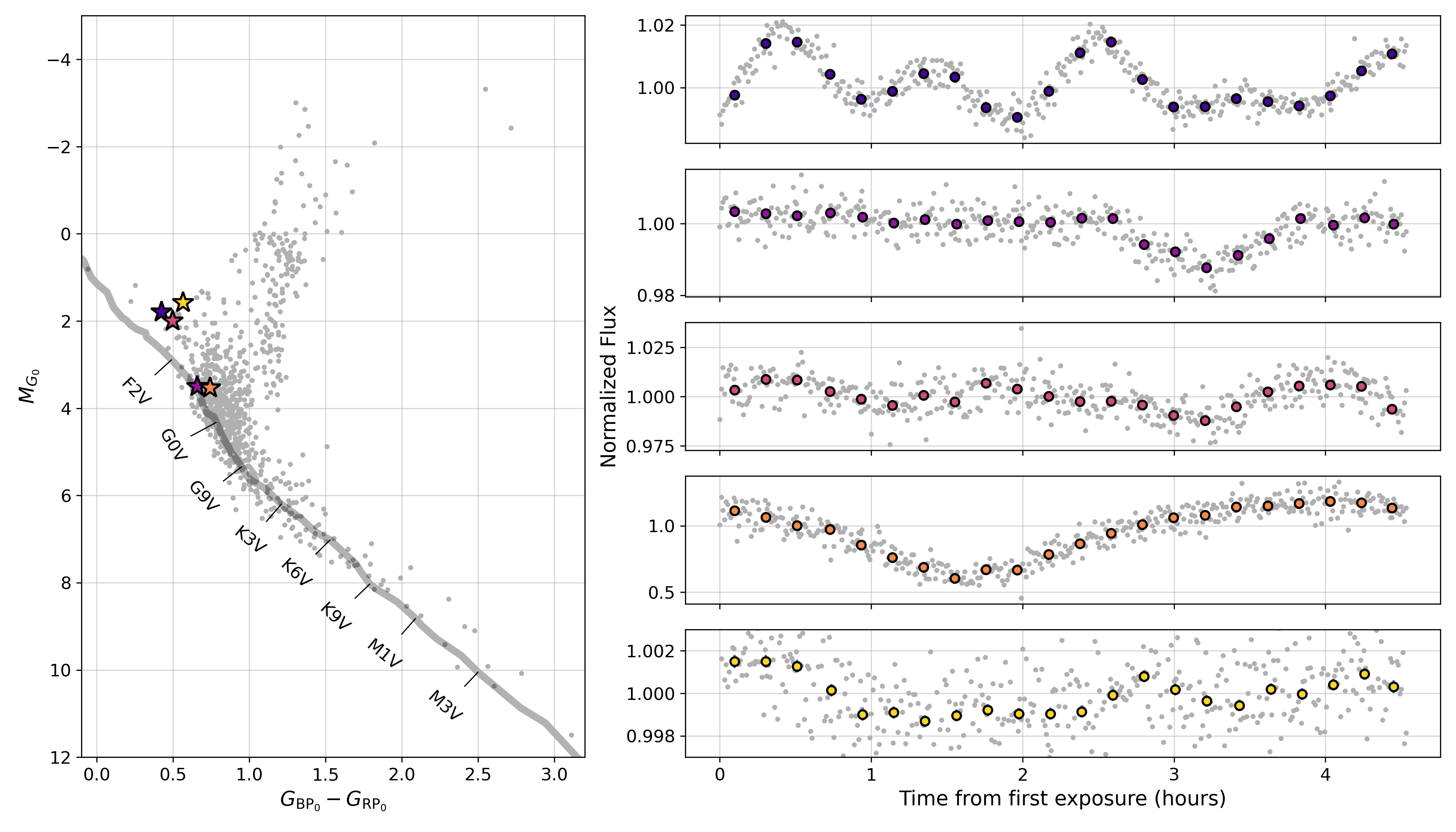}
    \caption{Extinction-corrected Gaia CMD (left) and \textit{Tierras} light curves (right) for several variable sources in the Kepler-1658 field. From top to bottom, these sources are: UCAC4~646-073016, a Delta Scuti variable star \citep{Balona2018}; KOI-6381, an eclipsing binary \citep{Conroy2014}; UCAC4~644-072117, a Delta Scuti variable star \citep{Murphy2019}; ATO~J294.3251+39.0326, an eclipsing binary \citep{Heinze2018}; and Kepler-1658, showing a ``transit" of Kepler-1658~b \citep{Chontos2019} \citep[actually the eclipse of a K/M binary diluted by an F-type primary;][]{Winn2025}. The predicted main sequence for the Gaia CMD from \cite{Pe13a} is indicated as a thick transparent line with select spectral type labels to guide the eye.}
    \label{fig:intranight_companion}
\end{figure*}

In Figure~\ref{fig:intranight_companion}, we present the Gaia color-magnitude diagram (CMD) for sources in the Kepler-1658 field.  We corrected the CMD for dust extinction, which dims and reddens starlight by wavelength dependent amounts. To recover intrinsic magnitudes, we first retrieved the native line-of-sight reddening $E$ from the 3D Milky Way dust map \texttt{Bayestar19} \citep{Green2019} as implemented in the Python package \texttt{dustmaps} \citep{Green2018} using the Gaia coordinate and distance \citep{BailerJones2021} of each source. We then convert this value into the standard Johnson $E(B-V)$ color excess and apply band-specific extinction corrections to each Gaia magnitude. 

The $E$ values from \texttt{Bayestar19} can be converted to $E(B-V)$ using 

\begin{align}
        E(g-r) &= 0.901E \label{eq:green}\\
        E(B-V) &= 0.98E(g-r) \label{eq:schalfly}\\
       \therefore E(B-V) &= 0.883 E
\end{align}
    
\noindent where Equation~\ref{eq:green} is from \cite{Green2019} and Equation~\ref{eq:schalfly} is from \cite{Schlafly2011}. The extinction-corrected Gaia magnitudes can then be calculated with

\begin{align}
        G_0 &= G-A_\text{G} = G - \frac{A_\text{G}}{A_\text{V}} R_\text{V}E(B-V)\\
        G_{\text{BP}_0} &= G_\text{BP} - A_{G_\text{BP}} = G_\text{BP} - \frac{A_{G_\text{BP}}}{A_\text{V}} R_\text{V}E(B-V) \\
        G_{\text{RP}_0} &=  G_\text{RP} - A_{G_\text{RP}} =  G_\text{RP} - \frac{A_{G_\text{RP}}}{A_\text{V}} R_\text{V}E(B-V)
\end{align}

\noindent where $\frac{A_\text{G}}{A_\text{V}} = 0.789$, $\frac{A_\text{BP}}{A_\text{V}} = 1.002$, $\frac{A_\text{RP}}{A_\text{V}} = 0.589$ \citep{Wang2019} and $R_\text{V} \equiv \frac{A_\text{V}}{E(B-V)} = 3.1$ \citep{Savage1979, Cardelli1989}.  

Figure~\ref{fig:intranight_companion} also shows light curves for several sources in the Kepler-1658 field on UT~2025~June~07. These sources were chosen because their light curves display significant variability as indicated in Figure~\ref{fig:intranight_precision}. They include two Delta-Scuti-type variable stars, two eclipsing binaries, and a ``transit" of Kepler-1658~b \citep{Chontos2019}, which was recently reclassified as an eclipsing K/M binary around an F-type primary \citep{Winn2025}. We note that the observations began at an airmass of 1.7 which likely caused the slight baseline slope seen in the transit of Kepler-1658~b.

As these results demonstrate, the photometric precision achieved by \textit{Tierras} on single observing nights very closely matches our theoretical expectations for the noise limit at a given source brightness. This, in combination with our pipeline which automatically creates light curves for all sources in the field, enables the identification and characterization of transient phenomena that occur on the timescales of hours, like variable stars, eclipsing binaries, and transiting exoplanets. 

\subsection{Night-to-Night Photometric Performance} \label{subsec:night_to_night}

Having demonstrated that \Tierras achieves intra-night precision consistent with theoretical expectations (Section~\ref{subsec:intranight_precision}), we now turn to the more challenging regime of night-to-night stability. As discussed in Section~\ref{sec:intro}, ground-based photometry has historically been unable to surpass a precision of $\sim$1~ppt on timescales longer than a few days \citep{Lockwood2007}, a barrier that limits the measurement of stellar rotation periods and other long-timescale phenomena central to the science goals outlined in Sections~\ref{sec:intro}. Two features of the \Tierras design are expected to mitigate the dominant contributors to this limit: the custom bandpass filter suppresses differential extinction from time-variable PWV (Section~\ref{sec:instrument}), and the wide field of view provides access to many reference stars whose combined artificial light curve reduces the impact of individual stellar variability on the ALC correction.

Beyond PWV, night-to-night precision is usually limited by flat-fielding errors, a systematic that MEarth overcame with two strategies that \Tierras inherited. First, our acquisition and guiding sequence (Section~\ref{subsec:computers}) places each target on the same detector pixel every night, accounting for the telescope pointing model and focal-plane distortion and using Gaia~DR3 astrometry. Each star therefore samples nearly the same pixels on every night, so any residual flat-field error acts as a near-constant factor that cancels in the normalized light curve. Second, the 60-day super flat (Section~\ref{subsec:flat_fields}) averages down both the photon noise and the night-to-night systematic variation of the flat-field correction.

In 2024, we performed a dedicated campaign with \textit{Tierras} to test the best-effort long-term precision of the data. We targeted a field containing LEP~1805-1422, one of the few active mid-M dwarfs within 15~pc that did not have a photometric rotation period measurement at the time \citep{Pass2023}. This source is located in a sight line near the galactic center ($l = 14.9^\circ; b = +3.3^\circ$) and \textit{Tierras} images of the field contain more than $4,000$ stars with $G_\text{RP} < 17$~mag., thus reducing the $\sigma_\text{ALC}/ALC$ term in Equation~\ref{eq:relative_sigma} compared to sparser fields. We observed the field with a 60-s exposure time for as long as it was observable on 19 nights spread over a time baseline of more than two months. We constructed light curves for all of the sources with the \textit{Tierras} pipeline and applied a quality mask to each, removing exposures with WCS solutions with an RMS larger than 0.215\arcsec (half a \textit{Tierras} pixel), $x$ or $y$ pointing deviations greater than 20~pixels, FWHM seeing values greater than 4\arcsec, and average reference star flux values less than 95\% of the median value (to eliminate exposures with cloud cover). We sigma-clipped the entire light curve with a clipping level of 4$\sigma$ and did the same to each night individually. 

\begin{figure*}
    \centering
    \includegraphics[width=\textwidth]{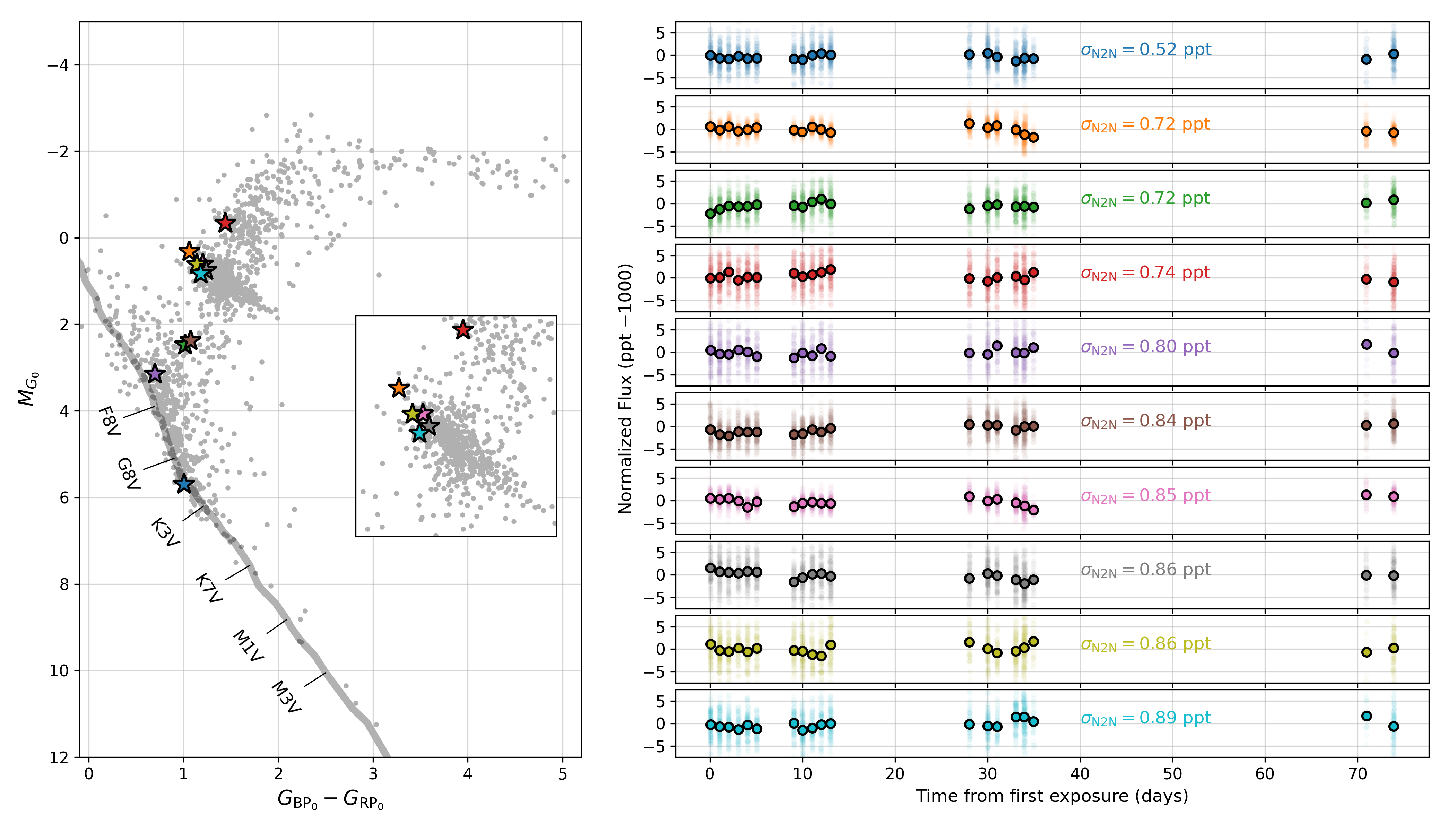}
    \caption{Left: Extinction-corrected Gaia CMD of sources in the LEP~1805-1422 field. The predicted main sequence from \cite{Pe13a} is indicated with a thick transparent line and select spectral types are indicated. The inset shows a zoom in on the red clump to make some of the highlighted points more visible. Right: The \textit{Tierras} light curves for the 10 stars with the lowest $\sigma_\text{N2N}$ values, all of which are less than 1~ppt over a time baseline of over 70 days. The light curves are shown in units of $\text{ppt} - 1000$ (on this scale, 5 corresponds to 0.5\%, for example). The transparent points show the photometry from individual exposures and the outlined points show the mean on each night. The colors of the starred objects in the left panel correspond to the light curve colors.}
    \label{fig:best_n2n}
\end{figure*}

We then calculated $\sigma_\text{N2N}$ for each source, which we define as the standard deviation of the set of nightly mean normalized flux values for each source. We present the 10 most stable light curves in Figure~\ref{fig:best_n2n}, along with their locations on an extinction-corrected Gaia CMD for the field. All of them have $\sigma_\text{N2N}$ less than 1~ppt, and our very best light curve has $\sigma_\text{N2N} = 0.52$~ppt. This demonstrates that the traditionally assumed 1-ppt noise floor of ground-based photometry is not an inherent physical limit and that purposeful instrument design and careful photometry can improve on it. In turn, this suggests a path toward achieving the long-term photometric precision that will be necessary to measure the rotation period of any Sun-like star. 

While a full investigation into the physical drivers of the light curve stability presented in Figure~\ref{fig:best_n2n} is beyond the scope of this work, we briefly note here that eight of the ten most stable sources appear to be evolved stars, with five sources in particular in the upper left part of the red clump (RC; see the inset of Figure~\ref{fig:best_n2n}). RC stars are evolved low-mass stars that are fusing He in their cores \citep[for a recent review, see ][]{Girardi2016}. They rotate slowly \citep[unless spun up by interaction with another object;][]{Tayar2015} and are observed to show little or no detectable surface rotation \citep{Ceillier2017, Tayar2018}. Sources near the top left of the RC in Gaia CMDs, in particular, are more metal-poor than other RC stars \citep{RuizDern2018}. For stars on the main sequence, lower metallicity implies shallower convective envelopes than high-metallicity stars at the same mass and rotation period, causing metal-poor stars to have higher Rossby numbers and correspondingly lower magnetic activity such as starspots which induce photometric variability \citep{Karoff2018, Amard2020, See2021, See2024}. Slow rotation and more homogeneous stellar surfaces would naturally explain the overdensity of extremely photometrically stable sources in the top-left of the RC in our dataset, but we leave the detailed mechanistic explanation of this phenomenon to future work. 

We note that we expect that we can demonstrate further improvements in our best-effort photometric stability with future observations. The LEP~1805-1422 field is near the southern declination limit for \textit{Tierras}, and thus all of the observations presented above were taken through a relatively high airmass (the average airmass of the observations was 1.5). The purposeful selection of a field near zenith would naturally reduce any airmass-dependent correlated noise and would also improve scintillation noise per Equation~\ref{eq:scint} (targeting a field at an average airmass of 1.1 would reduce the expected scintillation by a factor of $\left (\frac{1.5}{1.1} \right )^{7/4} = 1.7$, for example). Furthermore, the LEP~1805-1422 observations were taken before the optics upgrade described in Section~\ref{subsec:inner_baffles} and were therefore not flat fielded. Their sub-ppt stability despite this reflects the pixel-level pointing stability described above.

Finally, we wish to emphasize that the results shown in Figure~\ref{fig:best_n2n} represent our best-effort long-term photometric performance, which is only achievable if many exposures can be taken for a single field on each night. Standard \textit{Tierras} operations allocate far less time to any individual field on a given night. In particular, the ``standard" \textit{Tierras} observing mode breaks the night into 5-minute intervals and considers targets for each interval based on observability and proximity to the moon. For each interval, the scheduler selects the observable target for which the time between the last observation and the time of the interval is maximal\footnote{For new targets, the ``last" observation time is initialized to the start date of \textit{Tierras} observations, meaning that it will receive priority over all existing targets in the first interval when it becomes observable.}.  During actual observations, we require a minimum of three exposures to be taken during an interval, so that outliers due to, e.g., cosmic ray hits, can be easily discarded. 

With our standard observations, we observe typical $\sigma_\text{N2N}$ values that are around $1\textendash3$~ppt for the most stable stars in a field. While the standard observing mode provides slightly worse long-term precision than dedicated staring observations, it permits observations of many more fields, which we find to be desirable for achieving our scientific goals. Several \textit{Tierras}  publications have used such low-cadence observations to measure the rotation periods of exoplanet host stars \citep{Tamburo2025a, Yee2025, Tamburo2025b}, and we are using them to pursue rotation period measurements for volume-limited sample out to early-L spectral types (Tamburo et al. in prep.). 

\subsection{Observing Statistics}\label{subsec:obs_stats}
\begin{figure}
    \centering
    \includegraphics[width=\columnwidth]{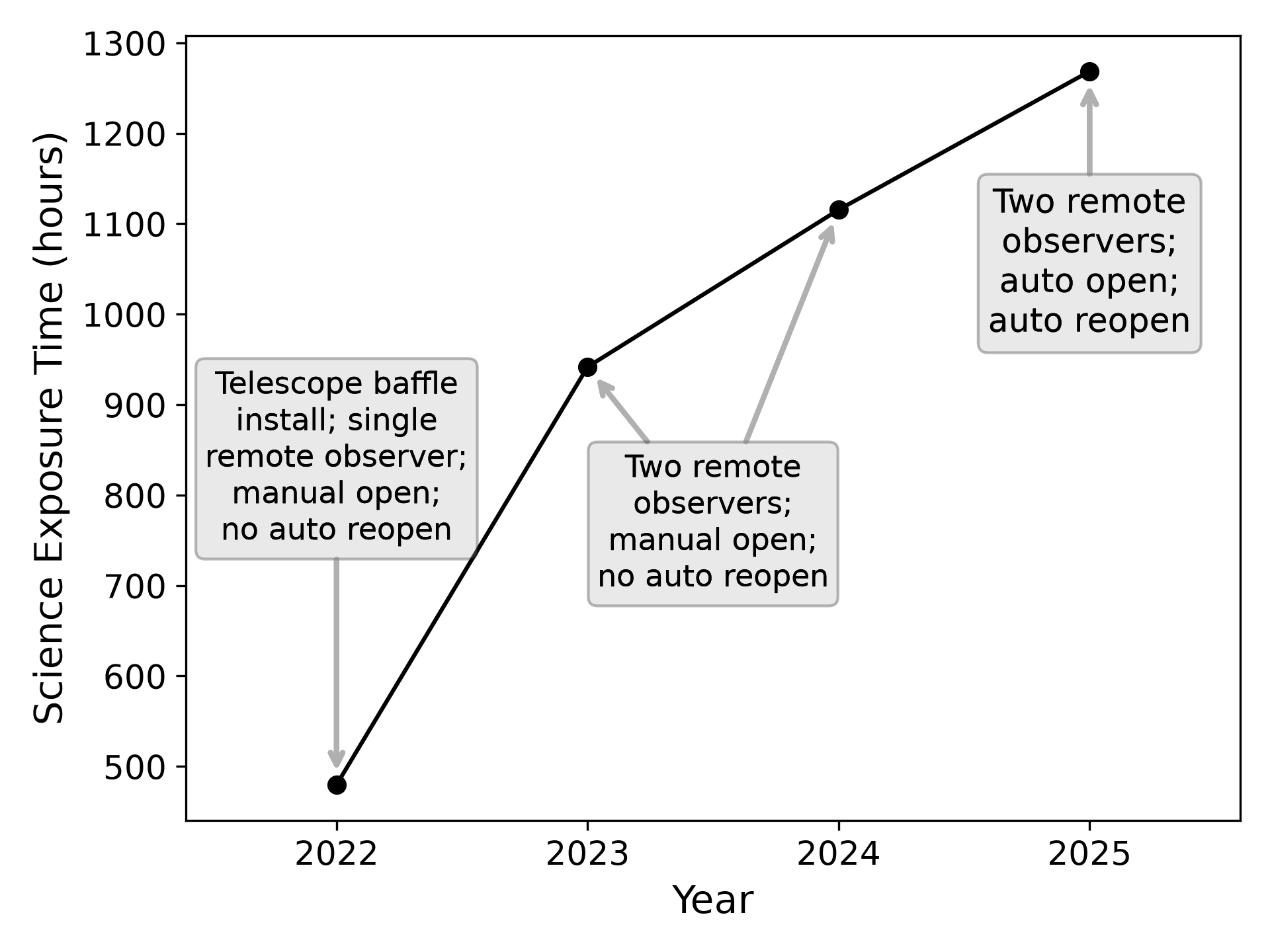}
    \caption{Total exposure time of science images obtained with \textit{Tierras} during each year since the project began full-time operation. An increasing amount of science exposure time has been attained each year due to progressively improving observing efficiency, as indicated in the text boxes.}
    \label{fig:exp_time_vs_year}
\end{figure}

Since achieving first light in late 2021, we have gradually improved the efficiency of observations with \textit{Tierras}. This is illustrated in Figure~\ref{fig:exp_time_vs_year}, which shows the cumulative exposure time of science images in each year since 2022, when nightly operations began. The amount of observing time per year has improved by more than 30 nights since the start of the project, largely thanks to a transition to more fully autonomous operations over time. For example, in 2022, we relied on a single telescope operator who had to issue remote commands to open the telescope, take flat fields, and initiate observing on each clear night. If the dome closed during the night due to weather flags being tripped, we had no mechanism to reopen the dome and restart observations outside of manual intervention in the middle of the night. In 2023, we began using two telescope operators, and in 2025 we upgraded the telescope software to automatically initiate observing on each clear night and to restart observations if they were interrupted and the weather later improved. Further gains can be made, for example by optimizing the observing schedule to minimize slew time. 

\begin{figure*}
    \centering
    \includegraphics[width=\textwidth]{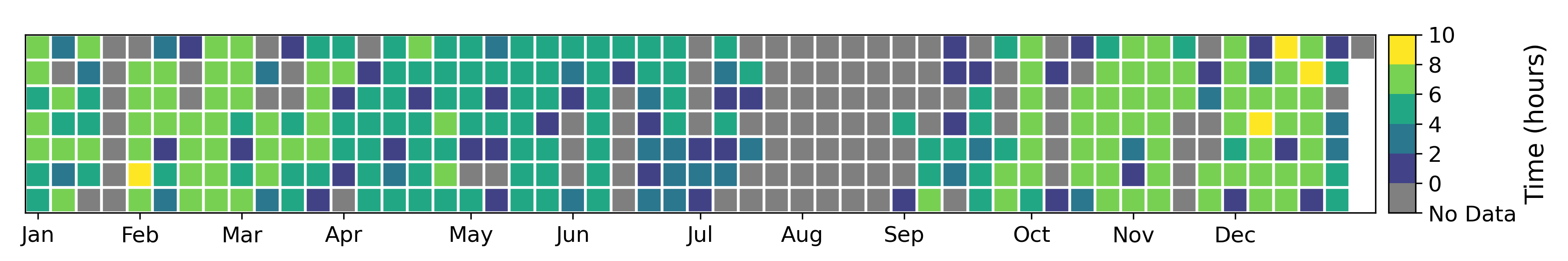}
    \caption{Total exposure time of science images for each day of the year in 2025.}
    \label{fig:on_sky_absolute}
\end{figure*}

In 2025, we obtained 1270~hours of science exposure time. In Figure~\ref{fig:on_sky_absolute}, we present a breakdown of how these data were gathered on each night of the year. The cells are colored by the amount of science time obtained on each night in hours and thus the change in the night length throughout the year is visually apparent. There is also a long series of nights where no data were obtained from late July to early September. This is due to the seasonal monsoon in Arizona, during which the observatory is closed yearly.

\begin{figure}
    \centering
    \includegraphics[width=\columnwidth]{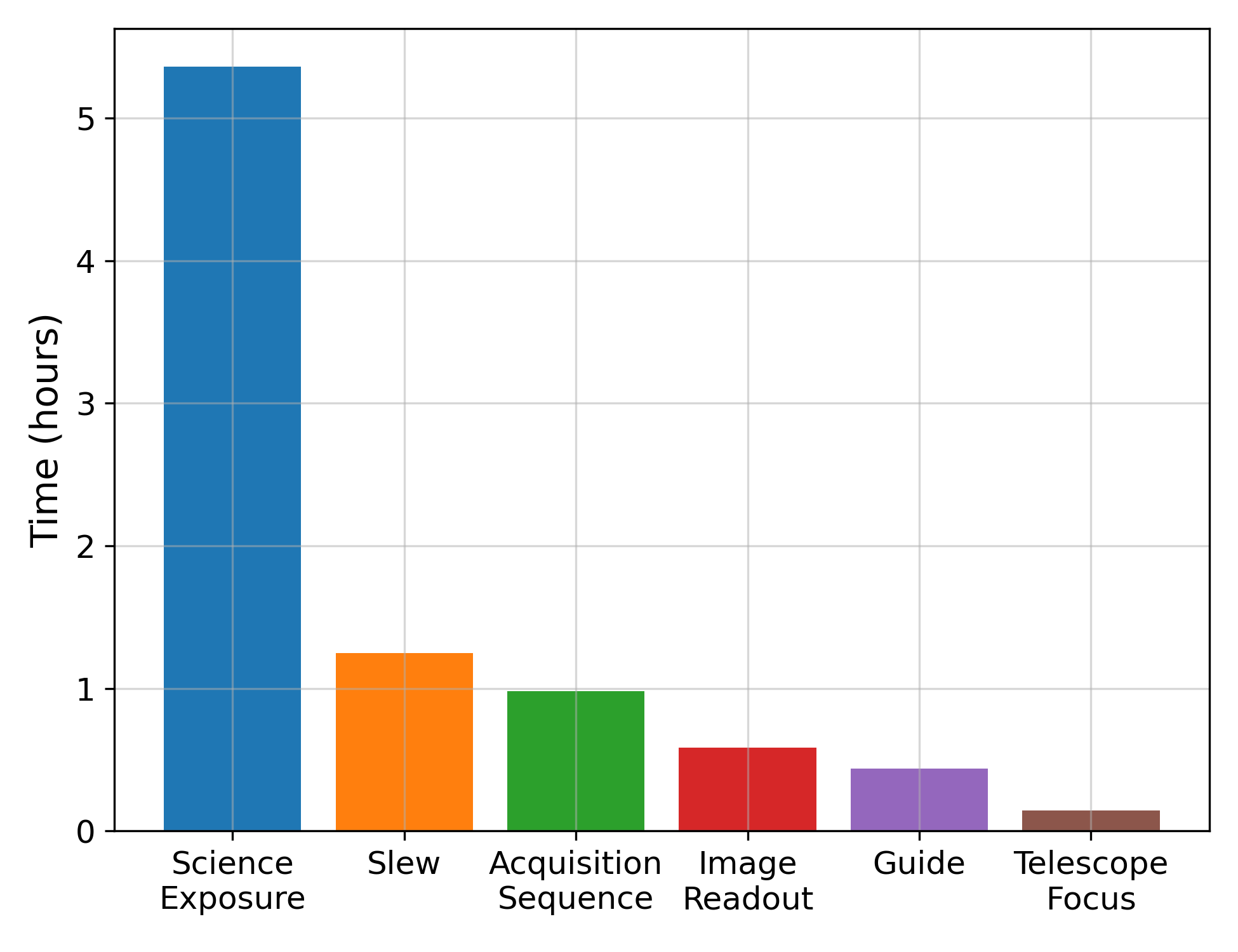}
    \caption{Breakdown of the median time per night spent on various observing activities across all nights with scientific data in 2025. See the text for a description of what each label entails.}
    \label{fig:avg_night_breakdown}
\end{figure}

Within a night, a variety of tasks are performed by the observing robot. In general, the observing sequence consists of (1) a slew to a new field. (2) an acquisition sequence in which 10-s exposures are taken and the telescope pointing corrected to place the target at the desired pixel position. (3) science exposure, (4) image readout. and (5) a guide correction to maintain the correct target pixel position. We also occasionally refocus the telescope throughout the night. We computed the median time spent on each of these tasks on every observing night in 2025 and present the results in Figure~\ref{fig:avg_night_breakdown}. The night-level observing efficiency for current \textit{Tierras} observations (the fraction of science exposure time to all observing tasks) is approximately 61\%, on average. This is a different quantity from the per-exposure duty cycle of Section~\ref{subsec:ccd}, which describes a single exposure cycle rather than a whole night. The non-science time is dominated by slews and acquisition sequences, each of which consume about an hour of time on a typical night. 

\section{Summary and Conclusions} \label{sec:summary}

The \Tierras Observatory is a fully robotic 1.3-m facility at FLWO designed from the ground up to reach sub-part-per-thousand precision on timescales of hours to months, with the latter being a regime that has eluded ground-based observatories for decades. Our design philosophy consisted of addressing all the dominant photometric noise sources architecturally rather than solely through post-observation corrections. The results we have presented in this paper demonstrate that our approach works: \Tierras is meeting its design goals and has produced the first ground-based light curves that break the 1~ppt night-to-night noise limit, demonstrating that this photometric precision barrier can be overcome through purpose-built instrumentation.

Our core design choices each target a specific term in the photometric 
error budget (Equation~\ref{eq:error_budget}). Our custom 40.2~nm FWHM bandpass filter centered at 863.5~nm suppresses PWV-induced errors to $\lesssim$0.25~ppt; an order of magnitude improvement over the long-pass 
filters used by other facilities in this line of work (Section~\ref{subsec:filter}). Our four-lens focal reducer and field-flattener expands the FOV to $0\fdg48 \times 0\fdg24$, providing the comparison star flux needed for robust differential photometry. Our deep-depletion CCD with 85\% QE in the \Tierras bandpass minimizes photon noise. Custom telescope baffles lined with IR flocking reduce the median sky background by ${\sim}42\times$ (Section~\ref{subsec:baffles}), and the entire system operates robotically with no nightly human intervention (Section~\ref{subsec:computers}).

Following installation in 2021 July, we commissioned the facility using a custom laser metrology system to align the primary and secondary mirrors, achieving a wavefront error of $\pm$15~nm (Section~\ref{subsec:collimation}). Prior to 2025 August, we had been unable to flat-field our images due to a large-scale radial flux gradient at the focal plane. In 2025 August, we installed lens-edge baffles inside the focal reducer assembly to suppress this scattered light, reducing the large-scale focal-plane flux gradient from ${\sim}10\%$ to ${\sim}2.5\%$ and enabling effective flat-fielding on \Tierras images for the first time (Section~\ref{subsec:inner_baffles}).

Our on-sky performance closely matches theoretical expectations. On single nights, the measured photometric scatter follows the predicted noise budget across the full range of source brightnesses, with several sources achieving better than 0.3~ppt precision in 20-minute bins (Section~\ref{subsec:intranight_precision}). To test our photometric performance limit on night-to-night timescales, we carried out a dedicated staring campaign on the LEP~1805-1422 field that yielded 10 light curves with $\sigma_\text{N2N} < 1$~ppt over a baseline exceeding 70~days, with the most stable reaching $\sigma_\text{N2N} = 0.52$~ppt (Section~\ref{subsec:night_to_night}). These are the first ground-based light curves to break the 1~ppt barrier on timescales of months, demonstrating that this limit can be overcome through purposeful instrument design and careful photometry. In standard survey mode, \Tierras achieves night-to-night precisions of 1--3~ppt for the brightest stars, sufficient to measure rotation periods and detect transiting planets across the many fields we monitor each season.

Our fully autonomous photometric pipeline processes \textit{all} sources in every observed field, typically producing updated light curves for thousands of stars on each observing night (Section~\ref{sec:pipeline}). This capability enables the systematic identification of variable stars, eclipsing binaries, and transiting exoplanets across the \Tierras survey footprint. We have already published several science results using \Tierras data, including rotation period measurements for the exoplanet host stars TOI-2364 \citep{Tamburo2025a}, TOI-2374 \citep{Yee2025}, and TOI-3884 \citep{Tamburo2025b}, as well as the detection of a transit of the long-period planet HIP~41378~f without observing ingress or egress \citep{GarciaMejia2026}. 

There is room for further improvement in our best-effort photometric stability. The observations presented in Section~\ref{subsec:night_to_night} were obtained through relatively high airmass and prior to the 2025 lens-edge baffle upgrade. We expect that targeting fields closer to zenith and incorporating flat-fielding will result in measurable gains. On the operations side, we are working to improve tracking performance through better modeling, including machine-learning methods (Zariski et al., in prep.). We are also working to reduce slew overhead through optimized target scheduling, and shorten our acquisition sequences. We are also exploring an upgrade that would allow integration and readout to overlap, so that the next exposure begins while the previous frame is read out. This would remove the 3.2~s readout from our per-exposure overhead and raise the duty cycle correspondingly. Finally, we aim to develop a public database to make our growing archive of light curves accessible to the broader stellar and exoplanet community.

These improvements will strengthen our scientific program and allow us to leverage \Tierras's unique long-term precision to 1) measure rotation periods for a volume-complete sample of nearby mid-to-late M dwarfs and brown dwarfs extending into the substellar regime; 2) characterize the orbits of the longest-period transiting exoplanets discovered by TESS and Kepler; 3) search for additional transiting terrestrial planets in known M dwarf systems; and 4) conduct a systematic search for exomoons and rings around longer-period planets. These efforts will inform the interpretation of JWST atmospheric observations and contribute to stellar characterization of future mission targets such as HWO's.

\begin{center}
    ACKNOWLEDGMENTS
\end{center}

The \textit{Tierras} Observatory is supported by the National Science Foundation under Award No.\ AST-2308043. The construction of \textit{Tierras} was made possible by grants from the John Templeton Foundation and the Harvard Origins of Life Initiative. The opinions expressed in this publication are those of the authors and do not necessarily reflect the views of the John Templeton Foundation. J.G.-M.\ acknowledges support from the Heising-Simons Foundation, and from the Pappalardo family through the MIT Pappalardo Fellowship in Physics.

\textit{Tierras} is located within the Fred Lawrence Whipple Observatory, and we are grateful to the FLWO staff, past and present, for their dedication to the maintenance of the mountain facilities and their enthusiastic support of this project. We especially thank Tom Gerl, Emilio Falco, and Wayne Peters for their contributions to the commissioning and operations of \textit{Tierras}. We thank Nelson Caldwell for his guidance on the design of the telescope baffles, Wade Poteet for guidance during the telescope refurbishment, and Christine Benoit for her tireless administrative support while building the observatory.

The authors would like to acknowledge Catherine Zucker for helpful conversations regarding the extinction correction of Gaia CMDs. 

We are grateful to our industry collaborators for their expertise and responsiveness: Gary Sims at Spectral Instruments; Frank Melsheimer, Ian Huss, and Damon Burke at DFM Engineering, Inc.; Alison Steves at Optimax; Chris Ghio at OHARA; Kevin Bennett, Alana Fustolo, and Joe Fustolo at Industrial Motions Engineering; Bill Carey at Atlantic Rubber; and Toshihiko Kimura at Asahi Spectra Co., Ltd.

This paper is dedicated to the memory of Gary (Scott) Sarapa, beloved member of the FLWO community, and Neil Pappalardo, whose generosity continues to support the next generation of physicists.

We used Claude (Anthropic) for three kinds of task in preparing this manuscript: analysis of archival \textit{Tierras} calibration and telemetry data underpinning the detector characterization reported in Section~\ref{subsec:ccd}, including the photon transfer curves, read noise, linearity, dark current, operating temperature distribution, and per-exposure duty cycle; generation of the left panel of Figure~\ref{fig:ccd_mosaic} and the bottom panel of Figure~\ref{fig:baffles}, and reformatting of Tables~\ref{tab:Telpresc} and \ref{tab:tierras_instrument}; and editorial feedback, including on the compliance of Figure~\ref{fig:architecture} with the C4 model convention, the notation in Section~\ref{sec:pipeline}, and an outline of Section~\ref{sec:summary}. Every number reported here was verified by the authors against the primary data or the vendor documentation, and the authors are fully responsible for the contents of this manuscript.

\facilities{\textit{Tierras} (FLWO 1.3m), GAIA} 

\software{\textit{dustmaps} \citep{Green2018}, \bedit{\textit{Astropy} \citep{astropy:2013,astropy:2018,astropy:2022}}, \bedit{\textit{numpy} \citep{numpy}}, \textit{pandas} \citep{pandas}, \textit{photutils} \citep{photutils}, \textit{scipy} \citep{scipy}, \textit{matplotlib} \citep{matplotlib}, \textit{Ansys Zemax OpticStudio}}

\bibliography{bib_file}{}
\bibliographystyle{aasjournal}

\end{document}